\documentclass[aps,prd,onecolumn,showpacs,showkeys,amsmath,amssymb]{revtex4}
\usepackage{epsfig} 
\usepackage{amsmath} 
\usepackage{makecell}
\usepackage{array}
\usepackage{subfigure}
\usepackage{bm}
\usepackage{longtable}
\usepackage{tabularx}
\usepackage{booktabs}
\usepackage{appendix}
\usepackage{caption2}
\usepackage{amsfonts}
\usepackage{amsmath}
\usepackage{graphicx}
\graphicspath{{figs/}}
\usepackage{subfigure}
\usepackage{dcolumn}
\usepackage{siunitx}
\usepackage{bm}
\usepackage{booktabs}
\usepackage[utf8]{inputenc}
\usepackage{float}
\usepackage{longtable,lscape}
\usepackage{txfonts}
\usepackage{overpic}
\usepackage{amssymb}
\usepackage{indentfirst}
\usepackage{epsfig}
\usepackage{feynmf}   
\usepackage{epstopdf}   
\usepackage{slashed}  
\usepackage{multirow}
\usepackage{natbib}

\usepackage[colorlinks, citecolor=blue,anchorcolor=red,menucolor=red, linkcolor=red,filecolor=red,runcolor=red,urlcolor=blue,frenchlinks=red]{hyperref}

\usepackage{ulem}
\usepackage{color}

\makeatletter
\newcommand{\figcaption}{\def\@captype{figure}\caption}
\newcommand{\tabcaption}{\def\@captype{table}\caption}

\newcommand{\Rmnum}[1]{\expandafter\@slowromancap\romannumeral #1@}

\def\hlinewd#1{%
  \noalign{\ifnum0=`}\fi\hrule \@height #1 \futurelet
   \reserved@a\@xhline}
\makeatother

\newcommand\dqq{\Big\langle \bar q q \Big\rangle}
\newcommand\dss{\Big\langle \bar s s \Big\rangle}
\newcommand\dqGq{\Big\langle g_s \bar q \sigma G q \Big\rangle}
\newcommand\dsGs{\Big\langle g_s \bar s \sigma G s \Big\rangle}
\newcommand\dGG{\Big\langle \alpha_{s} GG \Big\rangle}

\begin{document}
\title{Interpreting the $X(2370)$ and $X(2600)$ as light tetraquark states}

\author{Qi-Nan Wang$^{1,\,2}$}
\email{wangqinan@bhu.edu.cn}
\author{Ding-Kun Lian$^3$}
\email{liandk@mail2.sysu.edu.cn}

\author{Wei Chen$^{3,\, 4}$}
\email{chenwei29@mail.sysu.edu.cn}
\affiliation{$^1$College of Physical Science and Technology, Bohai University, Jinzhou 121013, China\\
$^2$Editorial Department of Journal, Bohai University, Jinzhou 121013, China\\ 
$^3$School of Physics, Sun Yat-Sen University, Guangzhou 510275, China\\
$^4$Southern Center for Nuclear-Science Theory (SCNT), Institute of Modern Physics, 
Chinese Academy of Sciences, Huizhou 516000, Guangdong Province, China}

\begin{abstract}
Inspired by the states $X(2370)$ and $X(2600)$ reported by the BESIII Collaboration, we systematically investigate the mass spectra of light compact tetraquark states with configurations $ud\bar{u}\bar{d}$, $us\bar{u}\bar{s}$, and $ss\bar{s}\bar{s}$ in the $J^{PC}=0^{-+}$ and $2^{-+}$ channels using the QCD sum rules approach. To effectively describe these tetraquark states, we construct appropriate interpolating tetraquark currents featuring three Lorentz indices while avoiding derivative operators. Through meticulous calculations of correlation functions up to dimension 10 condensates, we extract the mass spectra for both $0^{-+}$ and $2^{-+}$ states by employing the projection operator technique. Our results indicate that the masses of light tetraquarks span $1.5-2.5~\text{GeV}$ for $0^{-+}$ states and $2.4-2.7~\text{GeV}$ for $2^{-+}$ states. Notably, our analysis suggests that the $X(2370)$ state could be interpreted as a $0^{-+}$ $us\bar{u}\bar{s}$ or $ss\bar{s}\bar{s}$ tetraquark state, while the $X(2600)$ state is likely to be a $2^{-+}$ $us\bar{u}\bar{s}$ tetraquark. These intriguing findings warrant further detailed investigation in future studies to better understand the nature of these states.
\end{abstract}


\pacs{12.39.Mk, 12.38.Lg, 14.40.Ev, 14.40.Rt}
\keywords{Tetraquark, $X(2370)$, $X(2600)$, QCD sum rules}
\maketitle

\section{Introduction}
Quantum chromodynamics (QCD) serves as the foundational theory describing the strong interactions between quarks and gluons under the framework of SU(3) gauge symmetry. Beyond the conventional quark model, which encompasses two primary structures:mesons ($q\bar{q}$) and baryons ($qqq$)~\cite{Gell-Mann:1964ewy,Zweig:1964jf,ParticleDataGroup:2022pth}. QCD also predicts the existence of exotic hadronic states. These include tetraquarks, pentaquarks, hexaquarks, dibaryons, hybrid mesons, and glueballs~\cite{Meyer:2015eta,Chen:2016qju,Clement:2016vnl,Guo:2017jvc,Liu:2019zoy,Brambilla:2019esw,Chen:2022asf,Liu:2024uxn,Meng:2022ozq,Esposito:2016noz,Lebed:2016hpi,Wang:2025sic}.

To date, a plethora of new hadrons have been discovered in the light quark sector, including $\phi(2170)$~\cite{BaBar:2006gsq,BES:2007sqy,BESIII:2014ybv,BESIII:2017qkh,BaBar:2007ptr,BaBar:2007ceh,BaBar:2011btv,Belle:2008kuo}, $X(2239)$~\cite{BESIII:2018ldc}, and $\eta_{1}(1855)$~\cite{BESIII:2022iwi,BESIII:2022riz}, among others.
These experimental discoveries have spurred extensive theoretical investigations, with numerous interpretations proposed for these states, including the possibility of tetraquark configurations. For instance, the vector resonance $\phi(2175)$, with quantum numbers $J^{PC}=1^{--}$, has been considered a strong candidate for a fully strange $ss\bar{s}\bar{s}$ tetraquark state~\cite{Deng:2010zzd,Wang:2006ri,Ke:2018evd,Chen:2018kuu}, although alternative interpretations remain plausible~\cite{Ho:2019org,Deng:2013aca}. Similarly, the state $X(2239)$ has been interpreted as a P-wave $ss\bar{s}\bar{s}$ tetraquark with $J^{PC}=1^{--}$~\cite{Lu:2019ira,Azizi:2019ecm}. 
The exotic $1^{-+}$ channel has garnered significant attention, as its quantum numbers cannot be accommodated within the conventional quark model. Studies in ~\cite{Chen:2008ne,Chen:2008qw} have explored $I^{G}(J^{PC})=0^{+}(1^{-+})$ and $1^{-}(1^{-+})$ tetraquark states, yielding mass predictions in the ranges $1.7$--$2.1~\mathrm{GeV}$ and $1.6$--$2.0~\mathrm{GeV}$, respectively. These findings suggest that $\eta_{1}(1855)$ could be a tetraquark state, although alternative explanations, such as hybrid mesons~\cite{Chen:2022qpd,Qiu:2022ktc,Shastry:2022upd,Chen:2023ukh,Chen:2022isv,Shastry:2023ths} or molecular states~\cite{Wan:2022xkx,Yang:2022rck}, have also been proposed. Light tetraquark states have also been extensively studied across various quantum number channels, including exotic configurations such as $J^{PC}=0^{--}$~\cite{Jiao:2009ra,Huang:2016rro} and $0^{+-}$~\cite{Du:2012pn,Fu:2018ngx,Ray:2022fcl,Xi:2023byo}, as well as non-exotic channels like $J^{PC}=0^{++}$~\cite{Lodha:2024bwn}, $0^{-+}$~\cite{Su:2022eun,Lodha:2024bwn}, and $1^{+-}$~\cite{Wang:2019nln,Lodha:2024bwn,Cui:2019roq}. These studies continue to deepen our understanding of the complex spectrum of QCD bound states.

In 2010, the BESIII Collaboration reported a structure in $\pi \pi \eta^{\prime}$, named as $X(2370)$~\cite{BESIII:2010gmv}, which was confirmed in $K\bar{K}\eta^{\prime}$ channel~\cite{BESIII:2019wkp} in 2019. The mass and width  of $X(2370)$ are $2395 \pm 11_{-94}^{+26}~\mathrm{MeV}$ and $188_{-17-33}^{+18+124}~\mathrm{MeV}$ and its quantum numbers are determined as $J^{PC}=0^{-+}$~\cite{BESIII:2023wfi}. The discovery of this structure has attracted significant interest from researchers. It has been regarded as a glueball~\cite{Li:2025bko,Li:2024fko,Cao:2024mfn}, a $ss\bar{s}\bar{s}$ tetraquark state~\cite{Su:2022eun}, or the fourth radial excitation of $\eta(548)/\eta^{\prime}(958)$~\cite{Yu:2011ta}.
In 2022, the BESIII Collaboration reported a structure named $X(2600)$ with a statistical significance exceeding $20\sigma$ in the $\pi^{+}\pi^{-}\eta^{\prime}$ invariant mass spectrum. The mass and width of this state were measured to be $2618.3\pm2.0_{-1.4}^{+16.3}~\mathrm{MeV}$ and $195\pm5_{-17}^{+26}~\mathrm{MeV}$, respectively \cite{BESIIICollaboration:2022kwh}. This particle may possess quantum numbers of $0^{-+}$ or $2^{-+}$. In Ref.~\cite{Zhang:2022obn}, it was interpreted as a trigluon glueball with $J^{PC}=2^{-+}$ using the QCD sum rule method. However, in Ref.~\cite{Wang:2024yvo}, $X(2600)$ can also be viewed as the mesonic state $\eta_{2}^{\prime}(4D)$ based on a potential model. Further investigations into light tetraquark states were conducted in Refs.~\cite{Lodha:2024bwn,Lodha:2024yfn}, where the authors employed a potential model to estimate the masses of the ground $0^{-+}$ states, yielding values between $2.1-2.3~\mathrm{GeV}$ for a $\bar{\mathbf{3}}\otimes\mathbf{3}$ color configuration and $0.6-1.1~\mathrm{GeV}$ for a $\mathbf{6}\otimes \bar{\mathbf{6}}$ color configuration. For the ground $2^{-+}$ states, masses ranged from $2.4-2.7~\mathrm{GeV}$ in $\bar{\mathbf{3}}\otimes\mathbf{3}$ and $3.1-3.4~\mathrm{GeV}$ in $\mathbf{6}\otimes \bar{\mathbf{6}}$. The results obtained for the $\bar{\mathbf{3}}\otimes\mathbf{3}$ configuration align well with the masses of $X(2370)$ and $X(2600)$, suggesting that both $X(2370)$ and $X(2600)$ could be light tetraquark states.

To further investigate the inner structure of $X(2370)$ and $X(2600)$, this study aims to calculate the mass spectra of light tetraquark states with quantum numbers $J^{PC}=0^{-+},2^{-+}$, composed of various quark combinations such as $ud\bar{u}\bar{d}$, $us\bar{u}\bar{s}$, and $ss\bar{s}\bar{s}$, using the QCD sum rules method.

The paper is structured as follows: In Sec.~\ref{Sec:2}, we focus on constructing the interpolating tetraquark currents with three Lorentz indices suitable for coupling to physical states with $J^{PC}=0^{-+},2^{-+}$ and developing the projection operator to extract the invariant functions. Sec.~\ref{Sec:3} involves the calculation of correlation functions and spectral densities for these currents, leading to the establishment of mass sum rules for the tetraquark systems under consideration. Sec.~\ref{Sec:4} will present a numerical analysis to determine the tetraquark mass spectra. The final section will provide a concise summary and discussion of the results.

\section{Interpolating Currents and Projectors}\label{Sec:2}
In this section, we develop the interpolating currents for diquark-antidiquark pairs associated with light tetraquark states characterized by quantum numbers $J^{PC}=0^{-+},2^{-+}$. We explore various diquark fields such as $q_a^T C \gamma_{5}q_b$, $q_a^T C q_b$, $q_a^T C \gamma_\mu \gamma_{5}q_b$, $ q_a^T C \gamma_\mu q_b$, $ u_a^T C \sigma_{\mu \nu} q_b$, $q_a^T C \sigma_{\mu \nu}\gamma_5 q_b$, along with their corresponding antidiquark fields, where $a$ and $b$ represent color indices, and $T$ denotes the transpose of the matrices. The spin-parities of the diquark fields with various Lorentz structures are shown in Table ~\ref{Tab:diquark}. It should be noted that operators with one unit of orbital angular momentum ($L=1$, corresponding to $P$-wave) can be  constructed with or without covariant derivatives in the field operators. An example for the  former case is that $q_a^T C \gamma_5 \overleftrightarrow{D}_{\mu} q_b$ can couple to $^{1}P_{1}$ states. The latter case includes  the currents such as $q_{a}^{T}C q_{b}$, $q_{a}^{T}C\gamma_{i}\gamma_{5}q_{b}$, $q_{a}^{T}C\sigma_{ij}q_{b}$, $q_{a}^{T}C\sigma_{0i}\gamma_{5}q_{b}$ listed in Table ~\ref{Tab:diquark}, which carry $L=1$ despite containing no explicit derivatives. These currents are analogous to conventional $\bar{q}q$ (scalar mesons) and $\bar{q}\gamma_{\mu}\gamma_{5}q$ (axial-vector mesons), where the $P$-wave behavior originates from Lorentz structure and the charge conjugation properties.
One can construct a tetraquark operator as
\begin{equation}
O_{ij}= (q_a^{T}C\Gamma_{i} q_b)(\bar{q}_{a}\Gamma_{j} C \bar{q}_{b}^{T})\, ,
\end{equation}
where the color structures of the diquark and antidiquark fields depend on their Lorentz structures. It is easy to find the following identity under the charge conjugation transform~\cite{Chen:2010ze}
\begin{equation}
  \mathbb{C} O_{ij} \mathbb{C}^{-1}=O_{ij}^T\, .
\end{equation}

One can find the tetraquark operators with even and odd $C$-parities as 
\begin{equation}
    S=O_{ij}+O_{ij}^T\,, \quad A=O_{ij}-O_{ij}^T\,.
\end{equation}

\begin{table*}[t]
  \centering
    \caption{The spins and parities of the diquark fields.}
    \renewcommand\arraystretch{1.5}
  \setlength{\tabcolsep}{1.em}{
  \begin{tabular}{ccc}			\hline \hline
  $q^{T}C\Gamma q$   & $J^{P}$   & States   \\    \hline
   $q_{a}^{T}C\gamma_{5} q_{b}$    & $ 0^{+}$    & $^{1}S_{0}$    \\
   $q_{a}^{T}C q_{b}$      & $ 0^{-}$        & $^{3}P_{0}$     \\
    $q_{a}^{T}C\gamma_{\mu}q_{b}$  &  $ 1^{+}$&    $ ^{3}S_{1} $  \\
     $q_{a}^{T}C\gamma_{\mu}\gamma_{5}q_{b}$    & $\Big\{\begin{array}{rl}
                0^{+} & \mbox{if }\mu=0  \\
               1^{-} & \mbox{if }\mu=1,2,3
            \end{array} $      &  $\begin{array}{rl}
                ^{1}S_{0}  \\
                ^{3}P_{1}
            \end{array} $  \\
      $q_{a}^{T}C\sigma_{\mu\nu}q_{b}$    & $\Big\{\begin{array}{rl}
                1^{-} & \mbox{if }\mu,\nu=1,2,3  \\
               1^{+} & \mbox{if }\mu=0,\nu=1,2,3
            \end{array} $      &  $\begin{array}{rl}
                ^{1}P_{1}  \\
                ^{3}S_{1}
            \end{array} $  \\
   $q_{a}^{T}C\sigma_{\mu\nu}\gamma_{5}q_{b}$    & $\Big\{\begin{array}{rl}
                1^{+} & \mbox{if }  \mu,\nu=1,2,3\\
               1^{-} & \mbox{if }\mu=0,\nu=1,2,3 
            \end{array} $      &  $\begin{array}{rl} ^{3}S_{1}  \\
               ^{1}P_{1}
            \end{array} $    \\
            \hline \hline
  \label{Tab:diquark}
  \end{tabular}}
\end{table*}
In this work, we construct the following interpolating tetraquark currents with three Lorentz indices
\begin{equation}\label{Eq:current}
\begin{aligned}
    & J_{\alpha \mu \nu}^{1}=u_a^T C\gamma_{\alpha}  d_b\left(\bar{u}_a \sigma_{\mu \nu} C \bar{d}_b^T-\bar{u}_b \sigma_{\mu \nu} C \bar{d}_a^T\right) + u_a^T C \sigma_{\mu \nu} d_b\left(\bar{u}_a\gamma_{\alpha} C \bar{d}_b^T-\bar{u}_b\gamma_{\alpha} C \bar{d}_a^T\right)\,,\\
    & J_{\alpha \mu \nu}^{1\prime}=u_a^T C\gamma_{\alpha} d_b\left(\bar{u}_a \sigma_{\mu \nu} C \bar{d}_b^T+\bar{u}_b \sigma_{\mu \nu} C \bar{d}_a^T\right)  +  u_a^T C \sigma_{\mu \nu} d_b\left(\bar{u}_a\gamma_{\alpha} C \bar{d}_b^T+\bar{u}_b\gamma_{\alpha} C \bar{d}_a^T\right)\, , \\
    & J_{\alpha \mu \nu}^{2}=u_a^T C\gamma_{\alpha} \gamma_{5}d_b\left(\bar{u}_a \sigma_{\mu \nu} C \bar{d}_b^T-\bar{u}_b \sigma_{\mu \nu} C \bar{d}_a^T\right)  +  u_a^T C \sigma_{\mu \nu} d_b\left(\bar{u}_a\gamma_{\alpha} \gamma_{5}C \bar{d}_b^T-\bar{u}_b\gamma_{\alpha} \gamma_{5}C \bar{d}_a^T\right)\, ,\\
    & J_{\alpha \mu \nu}^{2\prime}=u_a^T C\gamma_{\alpha} \gamma_{5}d_b\left(\bar{u}_a \sigma_{\mu \nu} C \bar{d}_b^T+\bar{u}_b \sigma_{\mu \nu} C \bar{d}_a^T\right)  +  u_a^T C \sigma_{\mu \nu} d_b\left(\bar{u}_a\gamma_{\alpha} \gamma_{5}C \bar{d}_b^T+\bar{u}_b\gamma_{\alpha} \gamma_{5}C \bar{d}_a^T\right)\, ,\\
    & J_{\alpha \mu \nu}^{3}=u_a^T C\gamma_{\alpha} d_b\left(\bar{u}_a \sigma_{\mu \nu} \gamma_{5}C \bar{d}_b^T-\bar{u}_b \sigma_{\mu \nu} \gamma_{5}C \bar{d}_a^T\right)  +  u_a^T C \sigma_{\mu \nu} \gamma_{5}d_b\left(\bar{u}_a\gamma_{\alpha} C \bar{d}_b^T-\bar{u}_b\gamma_{\alpha} C \bar{d}_a^T\right)\, ,\\
    & J_{\alpha \mu \nu}^{3\prime}=u_a^T C\gamma_{\alpha} d_b\left(\bar{u}_a \sigma_{\mu \nu} \gamma_{5}C \bar{d}_b^T+\bar{u}_b \sigma_{\mu \nu} \gamma_{5}C \bar{d}_a^T\right)  +  u_a^T C \sigma_{\mu \nu} \gamma_{5}d_b\left(\bar{u}_a\gamma_{\alpha} C \bar{d}_b^T+\bar{u}_b\gamma_{\alpha} C \bar{d}_a^T\right)\, ,\\
    & J_{\alpha \mu \nu}^{4}=u_a^T C\gamma_{\alpha} \gamma_{5}d_b\left(\bar{u}_a \sigma_{\mu \nu} \gamma_{5}C \bar{d}_b^T-\bar{u}_b \sigma_{\mu \nu} \gamma_{5}C \bar{d}_a^T\right)  +  u_a^T C \sigma_{\mu \nu} \gamma_{5}d_b\left(\bar{u}_a\gamma_{\alpha} \gamma_{5}C \bar{d}_b^T-\bar{u}_b\gamma_{\alpha} \gamma_{5}C \bar{d}_a^T\right)\, ,\\
    & J_{\alpha \mu \nu}^{4\prime}=u_a^T C\gamma_{\alpha} \gamma_{5}d_b\left(\bar{u}_a \sigma_{\mu \nu} \gamma_{5}C \bar{d}_b^T+\bar{u}_b \sigma_{\mu \nu} \gamma_{5}C \bar{d}_a^T\right)  +  u_a^T C \sigma_{\mu \nu} \gamma_{5}d_b\left(\bar{u}_a\gamma_{\alpha} \gamma_{5}C \bar{d}_b^T+\bar{u}_b\gamma_{\alpha} \gamma_{5}C \bar{d}_a^T\right)\, , 
\end{aligned}
\end{equation}  
in which the currents $J_{\alpha \mu \nu}^{1}$, $J_{\alpha \mu \nu}^{2}$, $J_{\alpha \mu \nu}^{3}$ and $J_{\alpha \mu \nu}^{4}$ have the color structure  $\bar{\mathbf{3}}\otimes\mathbf{3}$, while $J_{\alpha \mu \nu}^{1\prime}$, $J_{\alpha \mu \nu}^{2\prime}$, $J_{\alpha \mu \nu}^{3\prime}$ and $J_{\alpha \mu \nu}^{4\prime}$ have the color structure $\mathbf{6}\otimes \bar{\mathbf{6}}$. 

For the fully strange $ss\bar s\bar s$ tetraquark systems, there are only two interpolating currents survived since the symmetric flavor structures
\begin{equation}
\begin{aligned}
     & J_{\alpha \mu \nu}^{ s1}=s_a^T C\gamma_{\alpha}  s_b\left(\bar{s}_a \sigma_{\mu \nu} C \bar{s}_b^T\right) + s_a^T C \sigma_{\mu \nu} s_b\left(\bar{s}_a\gamma_{\alpha} C \bar{s}_b^T\right)\, , \\
          & J_{\alpha \mu \nu}^{ s3}=s_a^T C\gamma_{\alpha}  s_b\left(\bar{s}_a \sigma_{\mu \nu}\gamma_{5}C \bar{s}_b^T\right) + s_a^T C \sigma_{\mu \nu}\gamma_{5} s_b\left(\bar{s}_a\gamma_{\alpha} C \bar{s}_b^T\right)\, ,
\end{aligned}  
\label{Eq:current_ssss}
\end{equation}
in which both of them have the antisymmetric color structure $\bar{\mathbf{3}}\otimes\mathbf{3}$.

In the investigation of tetraquark systems composed of a diquark and an antidiquark pair, the binding mechanism is primarily mediated by gluonic interactions. This short-range interaction leads to the formation of tightly bound states characterized by their compact spatial structure. Such systems can be adequately represented by local interpolating currents, mirroring the theoretical treatment applied to traditional hadronic states. People should note that two quark and two antiquark can also form a meson-meson molecular state and there is a certain correlation between the meson-meson current and the diquark-antiquark current. They can be transformed to each other through the Fierz rearrangement. However, generally speaking, a diquark-antidiquark current may couple to a physical tetraquark state more strongly than to a meson-meson molecular state, and hence it would be better to choose a diquark-antidiquark current to investigate tetraquark. However, the Fierz transformation is a purely mathematical operation that does not account for dynamical effects, whether perturbative or nonperturbative in nature. In contrast, meson-meson molecular configurations emerge as physical states generated through genuine hadronization processes. Moreover, the spatial structure of diquark-antidiquark systems typically exhibits significantly shorter binding distances than those of meson-meson molecular configurations. These substantial differences between the two systems may severely constrain the applicability of the Fierz transformation. To investigate the physical states with definite quantum numbers, we consider the  couplings between the interpolating current and different hadron states as follows
\begin{align}
\label{Eq:coupling1}
& \left\langle 0\left|J_{\alpha \mu \nu}\right| 0^{(-P) C}(p)\right\rangle=Z_1^0 p_\alpha g_{\mu \nu}+Z_2^0 p_\mu g_{\alpha \nu}+Z_3^0 p_\nu g_{\alpha \mu}+Z_4^0 p_\alpha p_\mu p_\nu  \,,\\ \label{Eq:coupling2}
& \left\langle 0\left|J_{\alpha \mu \nu}\right| 0^{PC}(p)\right\rangle=Z_5^0 \varepsilon_{\alpha \mu \nu \tau} p^\tau  \,,\\ \label{Eq:coupling3}
& \left\langle 0\left|J_{\alpha \mu \nu}\right| 1^{PC}(p)\right\rangle=Z_1^1 \epsilon_\alpha g_{\mu \nu}+Z_2^1 \epsilon_\mu g_{\alpha \nu}+Z_3^1 \epsilon_\nu g_{\alpha \mu}+Z_4^1 \epsilon_\alpha p_\mu p_\nu+Z_5^1 \epsilon_\mu p_\alpha p_\nu+Z_6^1 \epsilon_\nu p_\alpha p_\mu  \,,\\ \label{Eq:coupling4}
& \left\langle 0\left|J_{\alpha \mu \nu}\right| 1^{(-P) C}(p)\right\rangle=Z_7^1 \varepsilon_{\alpha \mu \nu \tau} \epsilon^\tau+Z_8^1 \varepsilon_{\alpha \mu \tau \lambda} \epsilon^\tau p^\lambda p_\nu+Z_9^1 \varepsilon_{\alpha \nu \tau \lambda} \epsilon^\tau p^\lambda p_\mu  \,,\\ \label{Eq:coupling5}
& \left\langle 0\left|J_{\alpha \mu \nu}\right| 2^{(-P) C}(p)\right\rangle=Z_1^2 \epsilon_{\alpha \mu} p_\nu+Z_2^2 \epsilon_{\alpha \nu} p_\mu+Z_3^2 \epsilon_{\mu \nu} p_\alpha  \,,\\ \label{Eq:coupling6}
& \left\langle 0\left|J_{\alpha \mu \nu}\right| 2^{PC}(p)\right\rangle=Z_4^2 \varepsilon_{\alpha \mu \tau \theta} \epsilon_\nu^{~\tau} p^\theta+Z_5^2 \varepsilon_{\alpha \nu \tau \theta} \epsilon_\mu^{~\tau} p^\theta  \,,\\
& \left\langle 0\left|J_{\alpha \mu \nu}\right| 3^{PC}(p)\right\rangle=Z_1^3 \epsilon_{\alpha \mu \nu} \, ,
\label{Eq:coupling7}
\end{align}
where $\epsilon_\alpha, \epsilon_{\alpha\mu}, \epsilon_{\alpha\mu\nu}$ are the polarization tensors for the spin-1, spin-2 and spin-3 states, respectively. $\varepsilon_{\alpha \mu \nu \tau}$ is the Levi-Civita tensor. The couplings in Eqs. (\ref{Eq:coupling1})-(\ref{Eq:coupling7}) represent the most general form of couplings for currents with three Lorentz indices and are independent of whether these indices exhibit symmetry or antisymmetry.
It should be noted that the interpolating currents in Eq.~(\ref{Eq:current}) and ~(\ref{Eq:current_ssss}) can not couple to any spin-3 state since their last two Lorentz indices are antisymmetric while the spin-3 polarization tensor $\epsilon_{\alpha\mu\nu}$ is completely symmetric.

Since the parities for the currents $J_{\alpha \mu \nu }^{1}$, $J_{\alpha \mu \nu }^{1\prime}$, $J_{\alpha \mu \nu }^{4}$, $J_{\alpha \mu \nu }^{4\prime}$, $J_{\alpha \mu \nu }^{s1}$ and $J_{\alpha \mu \nu }^{2}$, $J_{\alpha \mu \nu }^{2\prime}$, $J_{\alpha \mu \nu }^{3}$, $J_{\alpha \mu \nu }^{3\prime}$, $J_{\alpha \mu \nu }^{ s3}$ are opposite, they couple to the $0^{-+},2^{-+}$ tetraquark states via different coupling relations in Eq.~\eqref{Eq:coupling1}, Eq.~\eqref{Eq:coupling2}, Eq.~\eqref{Eq:coupling5}, and Eq.~\eqref{Eq:coupling6}, respectively. 
For the currents $J_{\alpha \mu \nu }^{1,1\prime,4,4\prime,s1}$, we can rewrite the coupling in Eq.~\eqref{Eq:coupling2} and \eqref{Eq:coupling6} in another way
\begin{equation}
  \begin{aligned}
  \left\langle 0\left|J_{\alpha \mu \nu}^{1,1\prime,4,4\prime, s1}\right| 0^{-+}(p)\right\rangle = f_{0-} \varepsilon_{\alpha \mu \nu \tau} p^\tau \, ,
  \end{aligned}
  \label{Eq:coupling2a}
\end{equation}
\begin{equation}
    \begin{aligned}
    \left\langle 0\left|J_{\alpha \mu \nu}^{1,1\prime,4,4\prime, s1}\right| 2^{-+}(p)\right\rangle &  =f_{2-}\left(\varepsilon_{\alpha \mu \tau \theta} \epsilon_\nu^{~\tau} p^\theta-\varepsilon_{\alpha \nu \tau \theta} \epsilon_\mu^{~\tau} p^\theta\right)\, ,
    \end{aligned}
    \label{Eq:coupling6a}
\end{equation}
in which $ f_{2-} $ represent the antisymmetric property under the exchange of $\mu$ and $\nu$. The symmetric part in Eq.~\eqref{Eq:coupling6} disappears due to the antisymmetric properties for all currents used in this work.
 
One can construct the normalized projection operators for the  $0^{-+},2^{-+}$ states
\begin{equation}
\mathbb{P}_{0}(\alpha_{1},\mu_{1},\nu_{1},\alpha_{2},\mu_{2},\nu_{2})=\frac{1}{36p^2}\varepsilon_{\alpha_{1} \mu_{1}  \nu_{1}  \tau_{1} } p^{\tau_{1}}
\varepsilon_{\alpha_{2}  \mu_{2} \nu_{2} \tau_{2}}  p^{\tau_{2}}  \,,
\label{Eq:Projector0a}
\end{equation}
\begin{equation}
  \mathbb{P}_{2}(\alpha_{1},\mu_{1},\nu_{1},\alpha_{2},\mu_{2},\nu_{2})=\frac{1}{20 p^2}\sum_{\lambda} \left(\varepsilon_{\alpha_{1} \mu_{1}  \tau_{1}  \theta_{1} } \epsilon_{\nu_{1} }^{~\tau_{1} }(p,\lambda) p^{\theta _{1}}-\varepsilon_{\alpha_{1}  \nu_{1}  \tau_{1}  \theta_{1} } \epsilon_{\mu_{1} }^{~\tau_{1} }(p,\lambda) p^{\theta_{1} }\right)
  \left(\varepsilon_{\alpha_{2}  \mu_{2} \tau_{2} \theta_{2}} \epsilon_{\nu_{2}}^{~\tau_{2}*}(p,\lambda) p^{\theta_{2}}-\varepsilon_{\alpha_{2} \nu_{2} \tau_{2} \theta_{2}} \epsilon_{\mu_{2}}^{~\tau_{2}*}(p,\lambda) p^{\theta_{2}}\right)   \,,
  \label{Eq:Projector2a}
  \end{equation}
where the summation over polarization $\lambda$ of the tensor $\epsilon_{\alpha \beta}(p,\lambda)$ is
\begin{equation}
\sum_{\lambda} \epsilon_{\alpha_{1} \beta_{1}}(p,\lambda)  \epsilon_{\alpha_{2} \beta_{2}}^{*}(p,\lambda) =\frac{1}{2}(\eta_{\alpha_{1} \alpha_{2}} \eta_{\beta_{1} \beta_{2}}+\eta_{\alpha_{1} \beta_{2}} \eta_{\beta_{1} \alpha_{2}}-\frac{2}{3} \eta_{\alpha_{1} \beta_{1}} \eta_{\alpha_{2} \beta_{2}}) \,,
\end{equation}
with
\begin{equation}
\eta_{\alpha\beta}=\frac{p_{\alpha} p_{\beta}}{p^{2}}-g_{\alpha \beta} \,.
\end{equation}
One might question the necessity of a coupling corresponding to the Lorentz structure $\varepsilon_{\mu \nu \tau \theta} \epsilon_\alpha^\tau p^\theta$ as indicated in Eq.~\eqref{Eq:coupling6}. However, it turns out that the projection projector formed by this structure is equivalent to that derived from the antisymmetric component in Eq.~\eqref{Eq:coupling6a}, resulting in only two independent tensor structures in Eq.~\eqref{Eq:coupling6}.

For the currents $J_{\alpha \mu \nu }^{2,2\prime,3,3\prime, s3}$, the coupling relation in Eq.~\eqref{Eq:coupling1} and \eqref{Eq:coupling5} can be rewritten as
\begin{equation}
  \begin{aligned}
\left\langle 0\left|J_{\alpha \mu \nu}^{2,2\prime,3,3\prime, s3}\right| 0^{-+}(p)\right\rangle &=f_{0-}^{\prime}( p_\mu g_{\alpha \nu}-p_\nu g_{\alpha \mu})\, ,
\end{aligned}
\label{Eq:coupling0b}
\end{equation}
\begin{equation}
  \begin{aligned}
\left\langle 0\left|J_{\alpha \mu \nu}^{2,2\prime,3,3\prime, s3}\right| 2^{-+}(p)\right\rangle &=f_{2-}^{\prime}(\epsilon_{\alpha \mu } p_\nu  -\epsilon_{\alpha \nu} p_\mu)\, ,
\end{aligned}
\label{Eq:coupling2b}
\end{equation}
with the normalized projection operators
\begin{equation}
  \mathbb{P}^{\prime}_{0}(\alpha_{1},\mu_{1},\nu_{1},\alpha_{2},\mu_{2},\nu_{2})=\frac{1}{36 p^2}( p_{\mu_{1}} g_{\alpha_{1} \nu_{1}}-p_{\nu_{1}} g_{\alpha_{1} \mu_{1}}) ( p_{\mu_{2}} g_{\alpha_{2} \nu_{2}}-p_{\nu_{2}} g_{\alpha_{2} \mu_{2}})   \, .
  \label{Eq:Projector0b}
\end{equation}
\begin{equation}
  \mathbb{P}^{\prime}_{2}(\alpha_{1},\mu_{1},\nu_{1},\alpha_{2},\mu_{2},\nu_{2})=\frac{1}{20 p^2}\sum_{\lambda} \left( \epsilon_{\alpha _{1} \mu_{1} }(p,\lambda) p_{\nu_{1}}-\epsilon_{\alpha _{1} \nu _{1} }(p,\lambda) p_{\mu _{1} }\right)
  \left( \epsilon_{\alpha _{2} \mu_{2} }^{\ast}(p,\lambda) p_{\nu_{2}}-\epsilon_{\alpha _{2} \nu _{2} }^{\ast}(p,\lambda) p_{\mu _{2} }\right) \, .
  \label{Eq:Projector2b}
\end{equation}
In our analysis, we observe that the $0^{-+}$ tetraquark states obtained from the currents $J_{\alpha \mu \nu}^{1, 1\prime, 2, 2\prime, s1}$ are just the same states as those derived from $J_{\alpha \mu \nu}^{3, 3\prime, 4, 4\prime, s3}$. Similar situations happen to $2^{-+}$ tetraquark states. Therefore, in the subsequent investigations, we focus solely on the interpolating currents $J_{\alpha \mu \nu}^{1}$, $J_{\alpha \mu \nu}^{1\prime}$, $J_{\alpha \mu \nu}^{2}$, $J_{\alpha \mu \nu}^{2\prime}$, and $J_{\alpha \mu \nu}^{s1}$ to explore the mass spectra of the $0^{-+}$ and $2^{-+}$ tetraquark states.

\section{Formalism of QCD sum rules}\label{Sec:3}
The two-point correlation functions of the currents in Eq.~(\ref{Eq:current}) and Eq.~(\ref{Eq:current_ssss}) can be written as
\begin{equation}
\begin{aligned}
 \Pi_{\alpha_{1}\mu_{1} \nu_{1},\alpha_{2}\mu_{2} \nu_{2}}(p^{2}) &=i \int d^{4} x e^{i p \cdot x}\left\langle 0\left|T\left[J_{\alpha_{1}\mu_{1} \nu_{1}}(x) J_{\alpha_{2}\mu_{2} \nu_{2}}^{\dagger}(0)\right]\right| 0\right\rangle \, .
 \label{Eq:correlator}
\end{aligned}
\end{equation}
We shall investigate the $J^{PC}=0^{-+},2^{-+}$ tetraquark states in this work, which can be extracted by applying the projection operators defined in Eq.~\eqref{Eq:Projector0a}, Eq.~\eqref{Eq:Projector0b}, Eq.~\eqref{Eq:Projector2a}, and Eq.~\eqref{Eq:Projector2b}
\begin{equation}
  \begin{aligned}
    \Pi(p^{2}) &=\mathbb{P}^{(\prime)}_{0,2}(\alpha_{1},\mu_{1},\nu_{1},\alpha_{2},\mu_{2},\nu_{2}) \Pi_{\alpha_{1}\mu_{1} \nu_{1},\alpha_{2}\mu_{2} \nu_{2}}(p^{2}) \, .
   \label{Eq:Pi_2}
  \end{aligned}
\end{equation}

At the hadronic level, the correlation function $\Pi(p^{2})$ can be usually described via the dispersion relation
\begin{equation}
\Pi(p^{2})=\frac{(p^{2})^{N}}{\pi} \int_{0}^{\infty} \frac{\operatorname{Im} \Pi(s)}{s^{N}\left(s-p^{2}-i \epsilon\right)} d s+\sum_{n=0}^{N-1} b_{n}(p^{2})^{n}\, ,
\label{Cor-Spe}
\end{equation}
where the $b_n$ is the subtraction constant. In QCD sum rules, the imaginary part of the correlation function is defined as the spectral function
\begin{equation}
\rho (s)\equiv\frac{1}{\pi} \text{Im}\Pi(s)=f^{2}m_{H}^{2}\delta(s-m_{H}^{2})+\text{QCD continuum and higher states}\, ,
\end{equation}
in which the “one pole plus continuum” parametrization assumption is used. The parameters $f$ and $m_{H}$ are the coupling constant and mass of the lowest-lying hadron state $H$, respectively. In QCD sum rule, this assumption has been widely used to study the properties of both conventional and exotic hadron states. The single pole in the spectral density originates from the zero-width approximation of the predicted hadron state. In reality, for tetraquark states, the dominant decay channels typically involve two-meson final states. This results in a coupling between the interpolating current and the two-meson continuum, thereby generating a finite width for the tetraquark state. However, the two-meson continuum effects can be incorporated through a renormalization of the coupling $f$ while preserving the tetraquark mass $m_{H} $, as discussed in Ref.~\cite{Wang:2020cme}. Since these width effects induce only minor corrections to the mass predictions, we adopt the zero-width single pole approximation in this work. However, one needs to require the pole dominance in the following analyses.

To enhance the convergence of the OPE series and to minimize the impact of the continuum and excited states, the Borel transformation technique is applied to the correlation functions at both the hadronic and quark-gluon levels. Subsequently, the QCD sum rules are derived as
\begin{equation}
\Pi\left(M_{B}^{2},s_{0}\right)=f^{2} m_{H}^{2}e^{-m_{H}^{2} / M_{B}^{2}}=\int_{0}^{s_{0}} d s e^{-s / M_{B}^{2}} \rho(s)\, ,
\end{equation}
where $M_{B}^{2}$ is the squared Borel mass introduced via the Borel transformation and $s_0$ is the continuum threshold.
Then the hadron mass of the lowest-lying tetraquark state can be extracted as
\begin{equation}
   \begin{aligned}
   m_{H}\left(M_{B}^{2},s_{0}\right) & =\sqrt{\frac{\frac{\partial}{\partial\left(-1 / M_B^2\right)} \Pi\left(M_{B}^{2},s_{0}\right)}{\Pi\left(M_{B}^{2},s_{0}\right)}} \,.
  \end{aligned}
\end{equation}

Since the Borel mass is an intermediate parameter, it should not be relevant to the physical state, i.e., to obtain the optimal value of the continuum threshold $s_0$, the variation of the extracted hadron mass $m$ with respect to $M_{B}^{2}$ should be minimized. The parameters $M_{B}^{2}$ and $s_{0}$ in the QCD sum rule analyses can be determined by requiring a suitable OPE convergence which in this work can be defined as
\begin{equation}
  R_{D>8}=\left|\frac{\Pi^{D>8}\left(M_{B}^{2}, \infty\right)}{\Pi^{t o t}\left(M_{B}^{2}, \infty\right)}\right| \,,
  \end{equation}
and a big enough pole contribution(P.C) which can also be defined as
\begin{equation}
  \text{P.C}=\frac{\Pi\left(M_{B}^{2}, s_0\right)}{\Pi\left(M_{B}^{2}, \infty\right)} \, .
\end{equation}
We calculate the correlation functions for tetraquark states with quantum numbers $J^{PC}=0^{-+},2^{-+}$, including condensates up to dimension 10. Contributions from operators of higher dimensions are negligible and are thus omitted. The detailed expressions, which are somewhat intricate, are provided in the Appendix. In the case of the nonstrange $ud\bar{u}\bar{d}$ configurations, the light quark masses are ignored in the chiral limit, precluding contributions from odd-dimensional condensates like quark and quark-gluon mixed condensates. One might suspect that neglecting the masses of the $u$ and $d$ quarks could have a significant impact on the results. However, our calculations show that this effect is extremely small, demonstrating that such a simplification is justified. For the $us\bar{u}\bar{s}$ and $ss\bar{s}\bar{s}$ systems, the mass of the strange quark is considered. For the high-dimensional condensates, we apply the factorization hypothesis and assume the associated factors to be 1. 

\section{Numerical analyses and mass predictions}\label{Sec:4}
In this section, we perform the QCD sum rule analyses for the light exotic tetraquark states with $J^{PC}=0^{-+},2^{-+}$. We use the following values for various QCD parameters~\cite{Jamin:2002ev,Narison:2011xe,Narison:2018dcr,ParticleDataGroup:2022pth}.
\begin{equation}
  \begin{aligned}
  m_{u} & = m_d=m_q=0\, ,\\
  m_{s} & =93_{-5}^{+11}  ~\mathrm{MeV}\, ,\\
  \dqq & =-(0.24 \pm 0.01)^3 ~\mathrm{GeV}^3 \,, \\
  \dss & =(0.8 \pm0.1)\times \dqq \,, \\
  \dqGq & =-(0.8  \pm 0.2) \times \dqq ~\mathrm{GeV}^2 \,, \\
  \dsGs & =(0.8 \pm 0.2) \times \dqGq\, ,\\ 
  \dGG & =(6.35  \pm 0.35) \times 10^{-2} ~\mathrm{GeV}^4 \,.
  \end{aligned}
\end{equation}
where $g_{s}$ is defined via $D_{\mu}=\partial_{\mu}+i g_{s}A_{\mu}$. The above condensate values are taken at energy renormalization scale $\mu=1\,\mathrm{GeV}$, and the $s$ quark mass is taken at $\mu=2\,\mathrm{GeV}$. To maintain energy renormalization scale consistency, we use renormalization group to choose the $s$ quark mass at  $\mu=1\,\mathrm{GeV}$.

\begin{figure}[t!!]
  \centering
  \includegraphics[width=8cm]{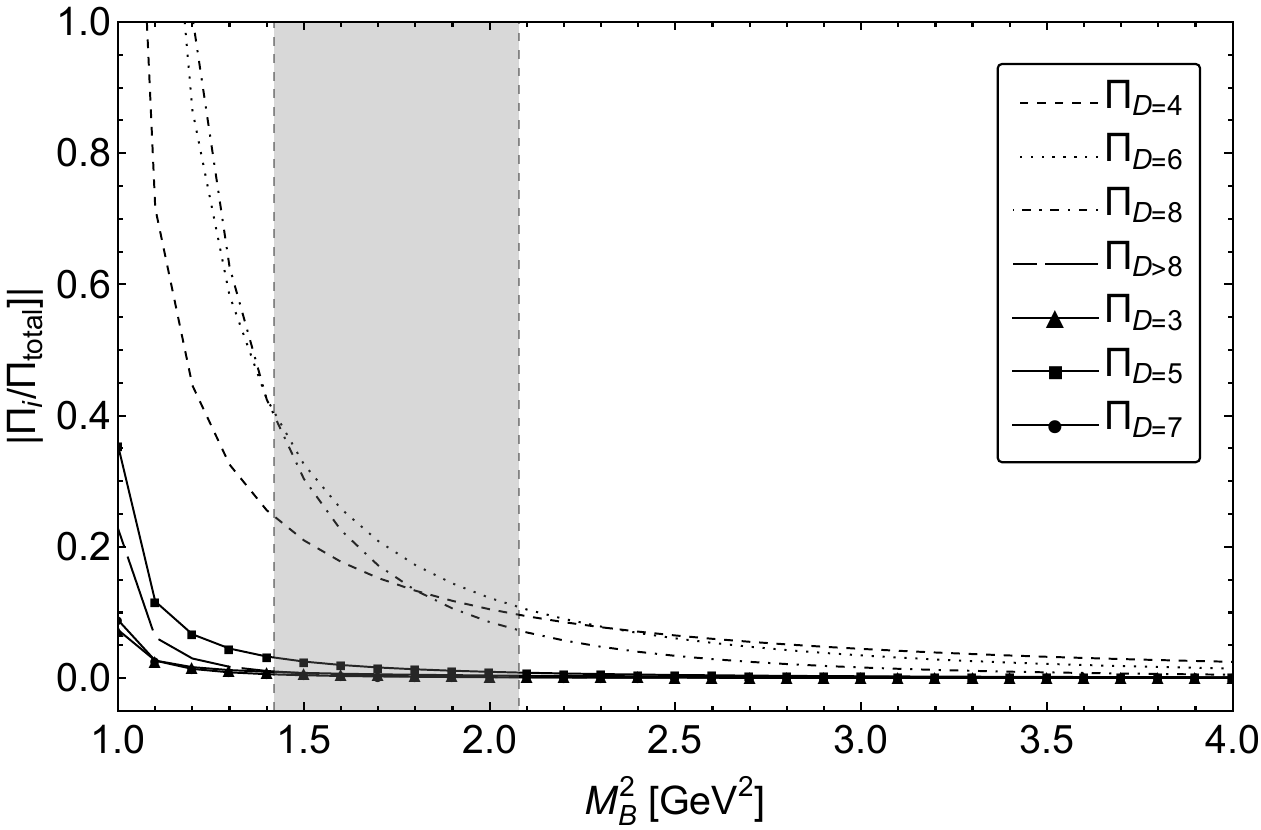}\\
  \caption{OPE convergence for the $us\bar{u}\bar{s}$ tetraquark state with $J^{PC}=2^{-+}$ extracted from the current $J_{\alpha\mu\nu}^{2}$.}
\label{fig:Convergence-s}
\end{figure}

In this study, we illustrate the numerical analysis focusing on the $2^{-+}$ $us\bar{u}\bar{s}$ tetraquark state from the current $J_{\alpha\mu\nu}^{2}$. To ensure the stability of the OPE series, it is imperative that the Borel parameter $M_{B}^{2}$ be sufficiently large. Our criterion is that the influence of condensates with dimensions higher than 8 should not exceed 1\% of the total contribution, establishing a minimum threshold for $M_{B}^{2}$ at $1.42~\mathrm{GeV}^{2}$. Fig.~\ref{fig:Convergence-s} displays the proportion of contributions from different condensates, confirming the satisfactory convergence of the OPE series.

\begin{figure}[t!!]
  \centering
  \subfigure[]{\includegraphics[width=8cm]{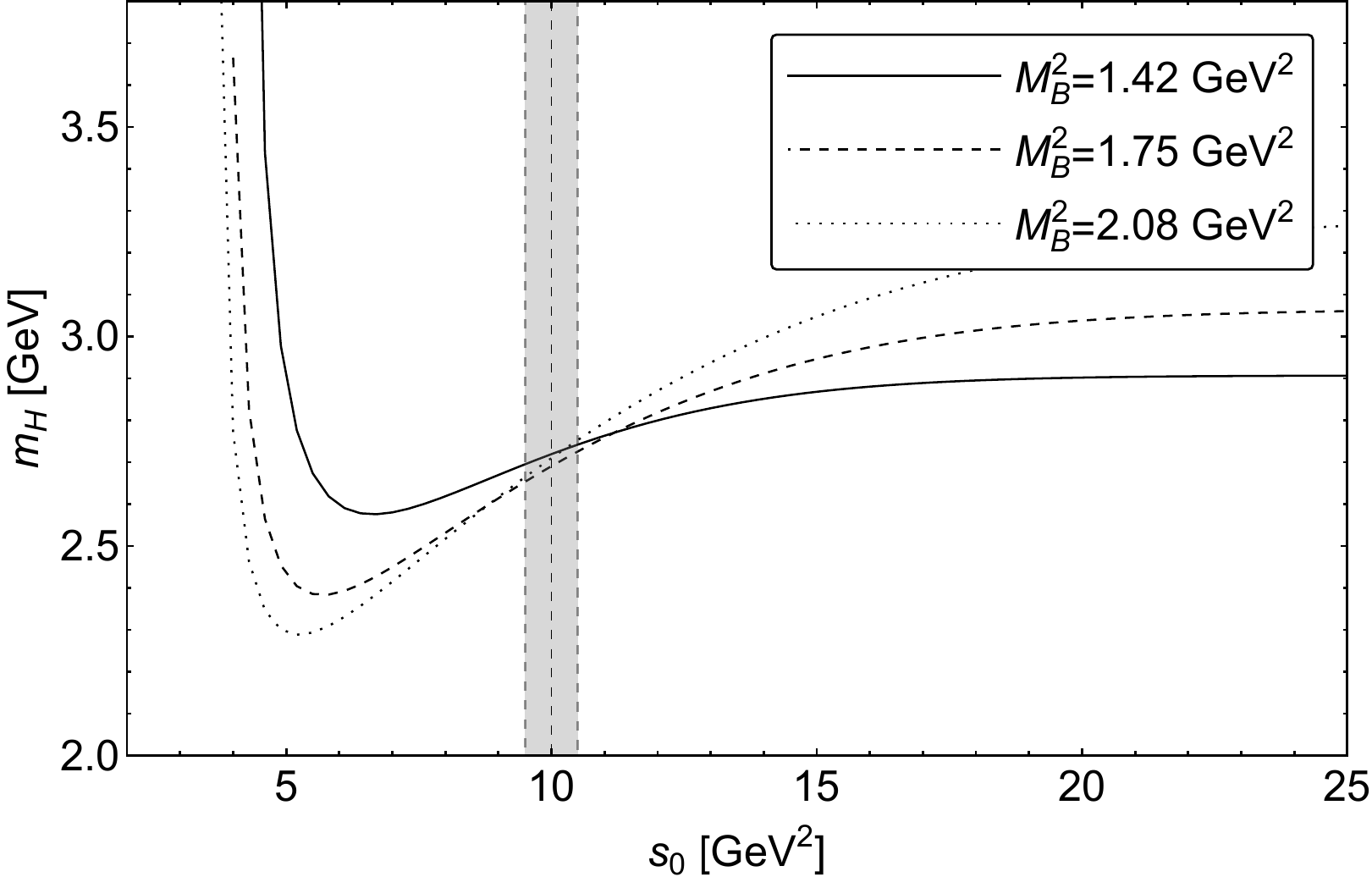}\label{fig:s0-mH-s}}
  \subfigure[]{\includegraphics[width=8.1cm]{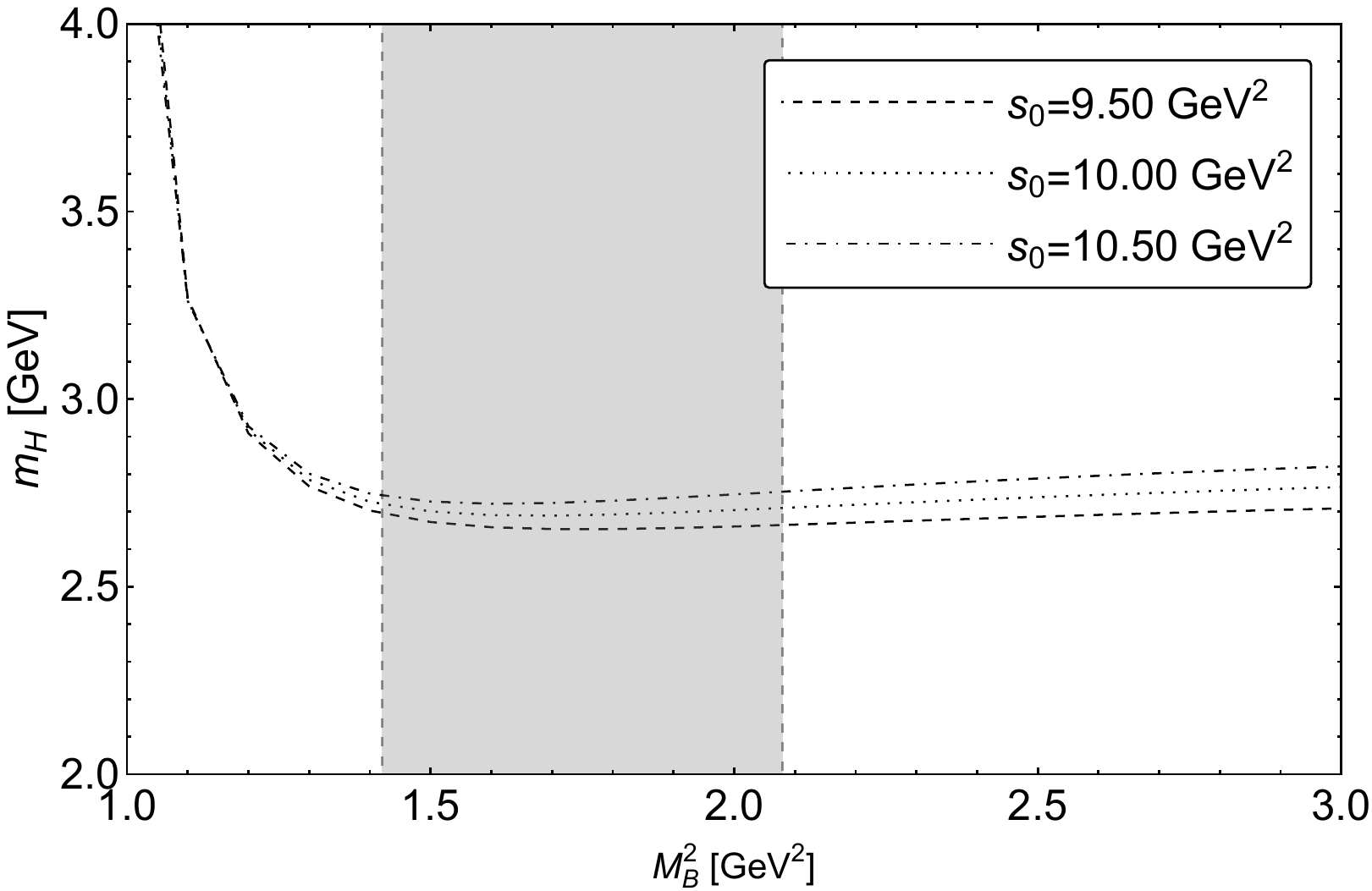}\label{fig:MB-mH-s}}\\
  \caption{Variation of $m_{H}$ with (a)$s_{0}$ and (b)$M_{B}^{2}$ corresponding to the $J^{PC}=2^{-+}$ $us\bar{u}\bar{s}$ tetraquark state extracted from current $J_{\alpha\mu\nu}^{2}$.}
\label{fig:Result-s}
\end{figure}

To determine the maximum value of $M_{B}^{2}$, the initial step is to establish the value of $s_{0}$. As discussed in Sec.~\ref{Sec:3}, the hadron mass $m_{H}$ should be independent of the Borel parameter $M_{B}^{2}$. Fig.~\ref{fig:s0-mH-s} depicts the relationship between $m_{H}$ and the continuum threshold $s_{0}$ for different values of $M_{B}^{2}$. The data indicate that the dependence of $m_{H}$ on $M_{B}^{2}$ is minimized at around $10.0~\mathrm{GeV}^{2}$. Outside the $s_{0}= 10~\mathrm{GeV}^{2}$ region, the predicted hadron masses exhibit strong dependence on the Borel parameter, while physical masses should remain independent of it. Thus, the working region for $s_{0}$ can be determined as $9.50\leq s_0\leq 10.50~\mathrm{GeV}^{2}$, where we assign a 5\% uncertainty to $s_{0}$. Subsequently, the upper limit for $M_{B}^{2}$ is established by ensuring that the P.C exceeds 50\%. Finally, the working region of the Borel parameter can be determined as $1.42\leq M_{B}^{2}\leq 2.08~\mathrm{GeV}^{2}$.

The leading nonperturbative effect in this system is attributed to the dimension 6 four-quark condensate in this region, while the dimension 4 and 8 condensate also play important roles. The contributions from the dimension 3 quark condensate and the dimension 5 quark-gluon mixed condensate have very small contributions due to the tiny mass of the strange quark. We present the Borel plots within the specified parameter regions in Fig.~\ref{fig:MB-mH-s}, where the stability of the QCD sum rules is adequate to reliably predict the hadron mass of the $2^{-+}$ $us\bar{u}\bar{s}$ tetraquark state
\begin{equation}
  \begin{aligned}
    m_{us\bar{u}\bar{s}}&=2.69_{-0.13}^{+0.11}~\mathrm{GeV}\, ,
  \end{aligned}
\end{equation}
where the errors are mainly from the uncertainties of the continuum threshold $s_{0}$, various QCD condensates. The error from the Borel mass is small enough to be neglected.

For all interpolating currents in Eq.~\eqref{Eq:current} and Eq.~\eqref{Eq:current_ssss}, similar numerical analysis can be performed. We collect the numerical results for the $ud\bar{u}\bar{d}$, $us\bar{u}\bar{s}$, and $ss\bar{s}\bar{s}$ tetraquark states with $J^{PC}=0^{-+}$, $2^{-+}$ in Table~\ref{tab:results_0mp} and \ref{tab:results_2mp}, respectively. It should be noted that the OPE convergence and P.Cs vary significantly for different channels. In our numerical analyses, we set the criteria of OPE convergence to be smaller than $1\%$, $5\%$, or $10\%$, and also set the criteria of P.C to be larger than $30\%$ or $50\%$ for different states from different currents. The larger the value of P.C or the smaller the value of $R_{D>8}$, the more reliable the predicted masses will be. Besides, we only get stable working regions of $M_{B}^{2}$ and $s_{0}$ for parts of the states from different currents. For the rest of the states, there are no working regions for $M_{B}^{2}$ and $s_{0}$, and thus no stable mass predictions can be made. We list these states as blanks in the tables.
\begin{center}
  \begin{table*}[t!]
    \caption{Mass spectra for $J^{PC}=0^{-+}$ light tetraquark states. The masses and P.Cs(values before ``$>$" in the last column) are evaluated at the central value of $s_{0}$ and $M_{B}^{2}$. The values in the fourth column are the criteria of OPE convergence. The values behind ``$>$" in the last column are the criteria of P.C.}
  \renewcommand\arraystretch{1.6} 
    \setlength{\tabcolsep}{1.6em}{
\begin{tabular}{ccccccc}
\hline\hline &Current  &$M_{B}^{2}(\mathrm{GeV}^{2})$ & $R_{D>8}$ & $s_{0}(\mathrm{GeV}^{2})$ & Mass$(\mathrm{GeV})$ &P.C\\ \hline
\hline \multirow{4}{*}{$u d \bar{u} \bar{d}$} 
 & $J_{\alpha \mu \nu}^{1}$  & $1.77-1.87$ & $<10\%$  & $6.0(\pm5\%)$ & $2.15_{-0.14}^{+0.16}$ & $32.0\%>30\%$ \\
 & $J_{\alpha \mu \nu}^{1\prime}$ & $-$ & $-$  & $-$ & $-$ & $-$ \\
 & $J_{\alpha \mu \nu}^{2}$ & $1.13-1.66$ & $<1\%$  & $5.3(\pm5\%$) & $1.95_{-0.07}^{+0.08}$ & $43.4\%>30\%$ \\
 & $J_{\alpha \mu \nu}^{2\prime}$ & $1.18-1.27$ & $<5\%$  & $3.0(\pm5\%)$ & $1.46_{-0.05}^{+0.07}$ & $32.5\%>30\%$ \\
 \cline{2-7}
  \multirow{4}{*}{$u s \bar{u} \bar{s}$}
 & $J_{\alpha \mu \nu}^{1}$ & $1.85-2.19$ & $<5\%$  & $7.3(\pm5\%)$ & $2.35_{-0.18}^{+0.19}$ & $35.9\%>30\%$ \\
 & $J_{\alpha \mu \nu}^{1\prime}$ &$1.55-1.90$ & $<5\%$  & $5.6(\pm5\%)$ & $2.04_{-0.19}^{+0.19}$ & $37.0\%>30\%$ \\
 & $J_{\alpha \mu \nu}^{2}$ & $1.42-1.98$ & $<5\%$  & $6.8(\pm5\%)$ & $2.24_{-0.17}^{+0.16}$ & $41.7\%>30\%$ \\
 & $J_{\alpha \mu \nu}^{2\prime}$ & $1.47-1.59$ & $<5\% $ & $4.5(\pm5\%)$ & $1.80_{-0.19}^{+0.19}$ & $33.1\%>30\%$ \\
 \cline{2-7} 
 $ss\bar{s}\bar{s}$ 
 & $J_{\alpha \mu \nu}^{s1}$ & $1.67-2.37$ & $<5\%$  & $8.1(\pm5\%)$ & $2.46_{-0.13}^{+0.14}$ & $42.3\%>30\%$ \\
\hline\hline
\end{tabular}
}
\label{tab:results_0mp}
\end{table*}
\end{center}
\begin{center}
  \begin{table*}[t!]
    \caption{Mass spectra for $J^{PC}=2^{-+}$ light tetraquark states. The masses and P.Cs(values before ``$>$" in the last column) are evaluated at the central value of $s_{0}$ and $M_{B}^{2}$. The values in the fourth column are the criteria of OPE convergence. The values behind ``$>$" in the last column are the criteria of P.C.}
  \renewcommand\arraystretch{1.6} 
    \setlength{\tabcolsep}{1.6em}{ 
\begin{tabular}{ccccccc}
\hline\hline &Current  &$M_{B}^{2}(\mathrm{GeV}^{2})$ & $R_{D>8}$ & $s_{0}(\mathrm{GeV}^{2})$ & Mass$(\mathrm{GeV})$ &P.C\\ \hline
\hline \multirow{4}{*}{$u d \bar{u} \bar{d}$} 
& $J_{\alpha \mu \nu}^{1}$ & $-$ & $-$  & $-$ & $-$ & $-$ \\
& $J_{\alpha \mu \nu}^{1\prime}$ & $-$ & $-$  & $-$ & $-$ & $-$ \\
& $J_{\alpha \mu \nu}^{2}$ & $1.53-2.06$ & $<5\%$  & $7.7(\pm5\%)$ & $2.43_{-0.08}^{+0.07}$ & $40.6\%>30\%$ \\
& $J_{\alpha \mu \nu}^{2\prime}$ & $1.75-1.92$ & $<5\%$  & $7.3(\pm5\%)$ & $2.40_{-0.09}^{+0.08}$ & $33.7\%>30\%$ \\
\cline{2-7} 
\multirow{4}{*}{$u s \bar{u} \bar{s}$} 
& $J_{\alpha \mu \nu}^{1}$ & $-$ & $-$  & $-$ & $-$ & $-$ \\
& $J_{\alpha \mu \nu}^{1\prime}$ & $-$ & $-$  & $-$ & $-$ & $-$ \\
& $J_{\alpha \mu \nu}^{2}$ & $1.42-2.08$ & $<1\%$  & $10.0(\pm5\%)$ & $2.69_{-0.13}^{+0.11}$ & $64.3\%>50\%$ \\
& $J_{\alpha \mu \nu}^{2\prime}$ & $1.52-2.00$ & $<5\%$  & $9.7(\pm5\%)$ & $2.68_{-0.13}^{+0.11}$ & $60.4\%>50\%$ \\
\cline{2-7} 
$ss\bar{s}\bar{s}$ 
& $J_{\alpha \mu \nu}^{s1}$ & $-$ & $-$ & $-$ & $-$ & $-$ \\
\hline\hline
\end{tabular}
}
\label{tab:results_2mp}
\end{table*}
\end{center}

\section{Conclusion and Discussion}\label{Sec:5}
In our study, we utilized QCD sum rules to examine the mass spectra of light tetraquark states, including $ud\bar{u}\bar{d}$, $us\bar{u}\bar{s}$, and $ss\bar{s}\bar{s}$, with quantum numbers $J^{PC}=0^{-+},2^{-+}$. This analysis involved constructing interpolating currents with three Lorentz indices. We computed the correlation functions and spectral densities, taking into account condensates up to dimension 10.

The anticipated masses for the tetraquark configurations with $J^{PC}=0^{-+}$ are estimated to be in the range of $1.5$ to $2.2~\mathrm{GeV}$ for $ud\bar{u}\bar{d}$, $1.8$ to $2.4~\mathrm{GeV}$ for $us\bar{u}\bar{s}$, and $2.46~\mathrm{GeV}$ for $ss\bar{s}\bar{s}$.
For the $2^{-+}$ tetraquark states, we obtain stable mass predictions only for $ud\bar{u}\bar{d}$ and $us\bar{u}\bar{s}$, with estimated masses around $2.4~\mathrm{GeV}$ and $2.7~\mathrm{GeV}$, respectively. Our predictions for the mass of $2^{-+}$ states are roughly consistent with those in Refs.~\cite{Lodha:2024bwn,Lodha:2024yfn} for the $\bar{\mathbf{3}}\otimes\mathbf{3}$ color configuration, but show significant discrepancies for the $\mathbf{6}\otimes\bar{\mathbf{6}}$ color structure.
Notably, the $0^{-+}$ $us\bar{u}\bar{s}$ state extracted from $J_{\alpha \mu \nu}^{1}$ and the $ss\bar{s}\bar{s}$ state derived from $J_{\alpha \mu \nu}^{s1}$ are in good agreement with the mass of $X(2370)$ in our predictions. This suggests that $X(2370)$ may be a $0^{-+}$ $us\bar{u}\bar{s}$ or $ss\bar{s}\bar{s}$ tetraquark state. Additionally, the mass of $X(2600)$ aligns well with our calculated mass for the $2^{-+}$ $us\bar{u}\bar{s}$ tetraquark state, indicating that $X(2600)$ could also be a $2^{-+}$ $us\bar{u}\bar{s}$ tetraquark state.

The isospin and $G$-parity values for the tetraquark systems vary: they are $I^{G}=0^{+}, 1^{-}, 2^{+}$ for the nonstrange $ud\bar{u}\bar{d}$ configuration, $I^{G}=0^{+}, 1^{-}$ for the $us\bar{u}\bar{s}$ setup, while the fully strange $ss\bar{s}\bar{s}$ system has $I^{G}=0^{+}$. Under SU(2) symmetry, the $u$ and $d$ quarks are treated equivalently, leading to the degeneracy of energy levels for tetraquark states with distinct isospins.
Since both $X(2370)$ and $X(2600)$ have been discovered in the invariant mass spectrum of $\pi \pi \eta^{\prime}$, and no clear isospin-2 particles have been found experimentally to date, it is likely that the $I^{G}$ of both particles is $0^{+}$. If this is the case, for $X(2370)$, the $f_{0}(980)\eta^{(\prime)}$ and $\phi h_{1}$ channels might be good choices to confirm its isospin and $G$-parity. Meanwhile, for $X(2600)$, the $\omega h_{1}$ and $\phi h_{1}$ channels could serve as suitable measurement channels.

It is anticipated that future theoretical and experimental research will delve deeper into these tetraquark states, as well as the $X(2370)$ and $X(2600)$.

\section*{ACKNOWLEDGMENTS}
This work is supported by the National Natural Science Foundation of China under Grant No. 12305147 and No. 12175318, the Natural Science Foundation of Guangdong Province of China under Grant No. 2022A1515011922, the Doctoral Startup Project of Bohai University under Grant No. 0525bs003.

\appendix
\section*{Appendix: Expressions of correlation functions}\label{appendix}
In this appendix, we show the expressions of correlation functions for the interpolating currents $J_{\alpha \mu \nu }^{1}$, $J_{\alpha \mu \nu }^{1\prime}$, $J_{\alpha \mu \nu }^{2}$, $J_{\alpha \mu \nu }^{2\prime}$ and $J_{\alpha \mu \nu }^{s1}$.
For the nonstrange $0^{-+}$ $ud\bar{u}\bar{d}$ tetraquark system, the correlation functions after Borel transformation are
\begin{equation}
\begin{aligned}
     \Pi_{d}^{1}(M_{B}^{2},s_{0})=&\int_{0}^{s_{0}} \left(
      \frac{s^4}{184320 \pi ^6}+\frac{\dGG s^2}{13824 \pi ^5}+\frac{\dqq^2 s}{36 \pi ^2}+\frac{23 \dqGq \dqq}{432 \pi ^2}
      \right)e^{-\frac{s}{M_{B}^{2}}} ds +
      \frac{7 \dqGq^2}{432 \pi ^2}+\frac{17 \dGG \dqq^2}{648 \pi }\, ,
\end{aligned}
\end{equation}

\begin{equation}
\begin{aligned}
       \Pi_{d}^{1'}(M_{B}^{2},s_{0})=&\int_{0}^{s_{0}} \left(\frac{s^4}{92160 \pi^6}+\frac{11 \dGG s^2}{13824 \pi^5}+\frac{\dqq^2 s}{18 \pi^2}+\frac{37 \dqGq \dqq}{432 \pi^2} \right)e^{-\frac{s}{M_{B}^{2}}} ds +
       \frac{19 \dqGq^2}{864 \pi^2}+\frac{25 \dGG \dqq^2}{648 \pi}\, ,
\end{aligned}
\end{equation}

\begin{equation}
\begin{aligned}
         \Pi_{d}^{2}(M_{B}^{2},s_{0})=&\int_{0}^{s_{0}} \left(
          \frac{s^4}{184320 \pi ^6}+\frac{\dGG s^2}{13824 \pi ^5}+\frac{\dqq^2 s}{108 \pi ^2}+\frac{\dqGq \dqq}{108 \pi ^2} \right)e^{-\frac{s}{M_{B}^{2}}} ds +
          \frac{\dqGq^2}{864 \pi ^2}-\frac{\dGG \dqq^2}{216 \pi }\, ,
\end{aligned}
\end{equation}

\begin{equation}
  \begin{aligned}
        \Pi_{d}^{2'}(M_{B}^{2},s_{0})=&\int_{0}^{s_{0}} \left(\frac{s^4}{92160 \pi ^6}+\frac{11 \dGG s^2}{13824 \pi ^5}+\frac{\dqq^2 s}{54 \pi ^2}+\frac{\dqGq \dqq}{216 \pi ^2}\right)e^{-\frac{s}{M_{B}^{2}}} ds
        -\frac{\dqGq^2}{216 \pi ^2}-\frac{\dGG \dqq^2}{216 \pi }\, .
\end{aligned}
\end{equation}

For the $0^{-+}$ $us\bar{u}\bar{s}$ tetraquark system, the correlation functions after Borel transformation are
\begin{equation}
\begin{aligned}
   \Pi_{s}^{1}(M_{B}^{2},s_{0})=&\int_{0}^{s_{0}} \left(
  \frac{s^4}{184320 \pi ^6}-\frac{\dqq m_{s} s^2}{1152 \pi ^4}-\frac{\dss m_{s} s^2}{1152 \pi ^4}+\frac{\dGG s^2}{13824 \pi ^5}+\frac{\dqq^2 s}{108 \pi ^2}+\frac{\dss^2 s}{108 \pi ^2}+\frac{\dqq \dss s}{108 \pi ^2}-\frac{19 \dqGq m_{s} s}{6912 \pi ^4}\right.\\ &\quad \left.-\frac{31 \dsGs m_{s} s}{6912 \pi ^4}+\frac{\dqGq \dqq}{64 \pi ^2}+\frac{\dsGs \dss}{64 \pi ^2}+\frac{19 \dqq \dsGs}{1728 \pi ^2}+\frac{19 \dqGq \dss}{1728 \pi ^2}\right.\\ &\quad \left.-\frac{29 \dGG \dqq m_{s}}{6912 \pi ^3}-\frac{17 \dGG \dss m_{s}}{6912 \pi ^3}
  \right)e^{-\frac{s}{M_{B}^{2}}} ds +
   \frac{5 \dqGq^2}{1152 \pi ^2}+\frac{13 \dsGs \dqGq}{1728 \pi ^2}\\ &\quad -\frac{\dGG m_{s} \dqGq}{768 \pi ^3}+\frac{\dqq \dss^2 m_{s}}{27} +\frac{2\dqq^2 \dss m_{s}}{9} +\frac{5 \dGG \dqq \dss}{324 \pi }+\frac{7 \dGG \dqq^2}{1296 \pi }\\ &\quad+\frac{7 \dGG \dss^2}{1296 \pi }+\frac{5 \dsGs^2}{1152 \pi ^2}-\frac{13 \dGG \dsGs m_{s}}{20736 \pi ^3}\, ,
\end{aligned}
\end{equation}

\begin{equation}
  \begin{aligned}
     \Pi_{s}^{1'}(M_{B}^{2},s_{0})=&\int_{0}^{s_{0}} \left(
      \frac{s^4}{92160 \pi ^6}-\frac{\dqq m_{s} s^2}{576 \pi ^4}-\frac{\dss m_{s} s^2}{576 \pi ^4}+\frac{11 \dGG s^2}{13824 \pi ^5}+\frac{\dqq^2 s}{54 \pi ^2}+\frac{\dss^2 s}{54 \pi ^2}+\frac{\dqq \dss s}{54 \pi ^2}-\frac{35 \dqGq m_{s} s}{6912 \pi ^4}\right.\\ &\quad \left.-\frac{83 \dsGs m_{s} s}{6912 \pi ^4}+\frac{13 \dqGq \dqq}{576 \pi ^2}+\frac{13 \dsGs \dss}{576 \pi ^2}+\frac{35 \dqq \dsGs}{1728 \pi ^2}+\frac{35 \dqGq \dss}{1728 \pi ^2}\right.\\ &\quad \left.-\frac{37 \dGG \dqq m_{s}}{6912 \pi ^3}-\frac{37 \dGG \dss m_{s}}{6912 \pi ^3}
      \right)e^{-\frac{s}{M_{B}^{2}}} ds+
      \frac{5 \dqGq^2}{1152 \pi ^2}+\frac{23 \dsGs \dqGq}{1728 \pi ^2}\\ &\quad-\frac{\dGG m_{s} \dqGq}{768 \pi ^3}+\frac{2\dqq \dss^2 m_{s}}{27} +\frac{4\dqq^2 \dss m_{s}}{9} +\frac{7 \dGG \dqq \dss}{324 \pi }+\frac{11 \dGG \dqq^2}{1296 \pi }\\ &\quad+\frac{11 \dGG \dss^2}{1296 \pi }+\frac{5 \dsGs^2}{1152 \pi ^2}-\frac{89 \dGG \dsGs m_{s}}{20736 \pi ^3}
      \, ,
\end{aligned}
\end{equation}

\begin{equation}
\begin{aligned}
    \Pi_{s}^{2}(M_{B}^{2},s_{0})=&\int_{0}^{s_{0}} \left(
    \frac{s^4}{184320 \pi ^6}+\frac{\dqq m_{s} s^2}{1152 \pi ^4}-\frac{\dss m_{s} s^2}{1152 \pi ^4}+\frac{\dGG s^2}{13824 \pi ^5}+\frac{\dqq^2 s}{108 \pi ^2}+\frac{\dss^2 s}{108 \pi ^2}-\frac{\dqq \dss s}{108 \pi ^2}+\frac{19 \dqGq m_{s} s}{6912 \pi ^4}\right.\\ &\quad \left.-\frac{31 \dsGs m_{s} s}{6912 \pi ^4}+\frac{\dqGq \dqq}{64 \pi ^2}+\frac{\dsGs \dss}{64 \pi ^2}-\frac{19 \dqq \dsGs}{1728 \pi ^2}-\frac{19 \dqGq \dss}{1728 \pi ^2}\right.\\ &\quad \left.+\frac{29 \dGG \dqq m_{s}}{6912 \pi ^3}-\frac{17 \dGG \dss m_{s}}{6912 \pi ^3}
    \right)e^{-\frac{s}{M_{B}^{2}}} ds+ 
    \frac{5 \dqGq^2}{1152 \pi ^2}-\frac{13 \dsGs \dqGq}{1728 \pi ^2}\\ &\quad+\frac{\dGG m_{s} \dqGq}{768 \pi ^3}-\frac{\dqq \dss^2 m_{s}}{27} +\frac{2 \dqq^2 \dss m_{s}}{9}-\frac{5 \dGG \dqq \dss}{324 \pi }+\frac{7 \dGG \dqq^2}{1296 \pi }\\ &\quad+\frac{7 \dGG \dss^2}{1296 \pi }+\frac{5 \dsGs^2}{1152 \pi ^2}-\frac{13 \dGG \dsGs m_{s}}{20736 \pi ^3}
    \, ,
\end{aligned}
\end{equation}
  
\begin{equation}
\begin{aligned}
    \Pi_{s}^{2'}(M_{B}^{2},s_{0})=&\int_{0}^{s_{0}} \left(
      \frac{s^4}{92160 \pi ^6}+\frac{\dqq m_{s} s^2}{576 \pi ^4}-\frac{\dss m_{s} s^2}{576 \pi ^4}+\frac{11 \dGG s^2}{13824 \pi ^5}+\frac{\dqq^2 s}{54 \pi ^2}+\frac{\dss^2 s}{54 \pi ^2}-\frac{\dqq \dss s}{54 \pi ^2}+\frac{35 \dqGq m_{s} s}{6912 \pi ^4}\right.\\ &\quad \left.-\frac{83 \dsGs m_{s} s}{6912 \pi ^4}+\frac{13 \dqGq \dqq}{576 \pi ^2}+\frac{13 \dsGs \dss}{576 \pi ^2}-\frac{35 \dqq \dsGs}{1728 \pi ^2}-\frac{35 \dqGq \dss}{1728 \pi ^2}\right.\\ &\quad \left.+\frac{37 \dGG \dqq m_{s}}{6912 \pi ^3}-\frac{37 \dGG \dss m_{s}}{6912 \pi ^3}
      \right)e^{-\frac{s}{M_{B}^{2}}} ds +
      \frac{5 \dqGq^2}{1152 \pi ^2}-\frac{23 \dsGs \dqGq}{1728 \pi ^2}\\ &\quad+\frac{\dGG m_{s} \dqGq}{768 \pi ^3}-\frac{2\dqq \dss^2 m_{s}}{27} +\frac{4\dqq^2 \dss m_{s}}{9} -\frac{7 \dGG \dqq \dss}{324 \pi }+\frac{11 \dGG \dqq^2}{1296 \pi }\\ &\quad+\frac{11 \dGG \dss^2}{1296 \pi }+\frac{5 \dsGs^2}{1152 \pi ^2}-\frac{89 \dGG \dsGs m_{s}}{20736 \pi ^3}
      \, .
\end{aligned}
\end{equation}
  
For the fully strange $0^{-+}$ $ss\bar{s}\bar{s}$ tetraquark system, the correlation functions after Borel transformation are
\begin{equation}
\begin{aligned}
    \Pi_{ss}^{s1}(M_{B}^{2},s_{0})=&\int_{0}^{s_{0}} \left(
      \frac{s^4}{184320 \pi ^6}-\frac{\dss m_{s} s^2}{288 \pi^4}+\frac{\dGG s^2}{13824 \pi ^5}+\frac{\dss^2 s}{36 \pi ^2}-\frac{25 \dsGs m_{s} s}{1728 \pi ^4}+\frac{23 \dsGs \dss}{432 \pi ^2}\right.\\ &\quad \left.-\frac{23 \dGG \dss m_{s}}{1728 \pi ^3}
      \right)e^{-\frac{s}{M_{B}^{2}}} ds +\frac{ 14 m_{s} \dss^3}{27}+\frac{17 \dGG \dss^2}{648 \pi }+\frac{7 \dsGs^2}{432 \pi ^2}-\frac{5 \dGG \dsGs m_{s}}{1296 \pi ^3}\, .
\end{aligned}
\end{equation}

For the nonstrange $2^{-+}$ $ud\bar{u}\bar{d}$ tetraquark system, the correlation functions after Borel transformation are
\begin{equation}
\begin{aligned}
       \Pi_{d}^{1}(M_{B}^{2},s_{0})=&\int_{0}^{s_{0}} \left(\frac{s^4}{53760 \pi ^6}-\frac{\dGG s^2}{2304 \pi ^5}-\frac{\dqq^2 s}{12 \pi ^2}-\frac{11 \dqGq \dqq}{72 \pi ^2}
        \right)e^{-\frac{s}{M_{B}^{2}}} ds-\frac{13 \dqGq^2}{288 \pi ^2}-\frac{\dGG \dqq^2}{54 \pi }\, ,
\end{aligned}
\end{equation}
  
\begin{equation}
\begin{aligned}
         \Pi_{d}^{1'}(M_{B}^{2},s_{0})=&\int_{0}^{s_{0}} \left(
          \frac{s^4}{26880 \pi ^6}-\frac{\dGG s^2}{1152 \pi ^5}-\frac{\dqq^2 s}{6 \pi ^2}-\frac{11 \dqGq \dqq}{36 \pi ^2} \right)e^{-\frac{s}{M_{B}^{2}}} ds-\frac{13 \dqGq^2}{144 \pi ^2}-\frac{\dGG \dqq^2}{27 \pi }\, ,
\end{aligned}
\end{equation}
  
\begin{equation}
\begin{aligned}
           \Pi_{d}^{2}(M_{B}^{2},s_{0})=&\int_{0}^{s_{0}} \left(
            \frac{s^4}{53760 \pi ^6}-\frac{\dGG s^2}{2304 \pi ^5}+\frac{\dqq^2 s}{36 \pi ^2}+\frac{7 \dqGq \dqq}{144 \pi ^2} \right)e^{-\frac{s}{M_{B}^{2}}} ds +
            \frac{\dqGq^2}{72 \pi ^2}\, ,
\end{aligned}
\end{equation}
  
\begin{equation}
\begin{aligned}
       \Pi_{d}^{2'}(M_{B}^{2},s_{0})=&\int_{0}^{s_{0}} \left(\frac{s^4}{26880 \pi ^6}-\frac{\dGG s^2}{1152 \pi ^5}+\frac{\dqq^2 s}{18 \pi ^2}+\frac{17 \dqGq \dqq}{144 \pi ^2}\right)e^{-\frac{s}{M_{B}^{2}}} ds +
       \frac{11 \dqGq^2}{288 \pi ^2}+\frac{\dGG \dqq^2}{18 \pi }\, .
\end{aligned}
\end{equation}

For the $2^{-+}$ $us\bar{u}\bar{s}$ tetraquark system, the correlation functions after Borel transformation are
\begin{equation}
\begin{aligned}
      \Pi_{s}^{1}(M_{B}^{2},s_{0})=&\int_{0}^{s_{0}} \left(
      \frac{s^4}{53760 \pi ^6}+\frac{\dqq m_{s} s^2}{192 \pi ^4}+\frac{13 \dss m_{s} s^2}{1920 \pi ^4}-\frac{\dGG s^2}{2304 \pi ^5}-\frac{\dqq \dss s}{18 \pi ^2}-\frac{\dqq^2 s}{72 \pi ^2}-\frac{\dss^2 s}{72 \pi ^2}+\frac{29 \dqGq m_{s} s}{2304 \pi ^4}\right.\\ &\quad \left.+\frac{29 \dsGs m_{s} s}{2304 \pi ^4}-\frac{5 \dqGq \dqq}{192 \pi ^2}-\frac{5 \dsGs \dss}{192 \pi ^2}-\frac{29 \dqq \dsGs}{576 \pi ^2}-\frac{29 \dqGq \dss}{576 \pi ^2}\right.\\ &\quad \left.-\frac{5 \dGG \dqq m_{s}}{2304 \pi ^3}+\frac{\dGG \dss m_{s}}{2304 \pi ^3}
      \right)e^{-\frac{s}{M_{B}^{2}}} ds 
      -\frac{\dqGq^2}{128 \pi ^2}-\frac{17 \dsGs \dqGq}{576 \pi ^2}-\frac{\dGG m_{s} \dqGq}{256 \pi ^3}\\ &\quad-\frac{2\dqq \dss^2 m_{s}}{9} +\frac{\dqq^2 \dss m_{s}}{3} -\frac{\dGG \dqq \dss}{108 \pi }-\frac{\dGG \dqq^2}{216 \pi }-\frac{\dGG \dss^2}{216 \pi }-\frac{\dsGs^2}{128 \pi ^2}\\ &\quad-\frac{\dGG \dsGs m_{s}}{6912 \pi ^3} \, ,
\end{aligned}
\end{equation}

\begin{equation}
\begin{aligned}
     \Pi_{s}^{1'}(M_{B}^{2},s_{0})=&\int_{0}^{s_{0}} \left(
   \frac{s^4}{26880 \pi ^6}+\frac{\dqq m_{s} s^2}{96 \pi ^4}+\frac{13 \dss m_{s} s^2}{960 \pi ^4}-\frac{\dGG s^2}{1152 \pi ^5}-\frac{\dqq \dss s}{9 \pi ^2}-\frac{\dqq^2 s}{36 \pi ^2}-\frac{\dss^2 s}{36 \pi ^2}+\frac{61 \dqGq m_{s} s}{2304 \pi ^4}\right.\\ &\quad \left.+\frac{55 \dsGs m_{s} s}{2304 \pi ^4}-\frac{3 \dqGq \dqq}{64 \pi ^2}-\frac{3 \dsGs \dss}{64 \pi ^2}-\frac{61 \dqq \dsGs}{576 \pi ^2}-\frac{61 \dqGq \dss}{576 \pi ^2}\right.\\ &\quad \left.+\frac{11 \dGG \dqq m_{s}}{2304 \pi ^3}-\frac{19 \dGG \dss m_{s}}{2304 \pi ^3}
   \right)e^{-\frac{s}{M_{B}^{2}}} ds 
   -\frac{5 \dqGq^2}{384 \pi ^2}-\frac{37 \dsGs \dqGq}{576 \pi ^2}\\ &\quad-\frac{\dGG m_{s} \dqGq}{256 \pi ^3}-\frac{4\dqq \dss^2 m_{s}}{9} +\frac{2\dqq^2 \dss m_{s}}{3} -\frac{5 \dGG \dqq \dss}{108 \pi }+\frac{\dGG \dqq^2}{216 \pi }\\ &\quad+\frac{\dGG \dss^2}{216 \pi }-\frac{5 \dsGs^2}{384 \pi ^2}-\frac{29 \dGG \dsGs m_{s}}{6912 \pi ^3}
   \, ,
\end{aligned}
\end{equation}

\begin{equation}
\begin{aligned}
     \Pi_{s}^{2}(M_{B}^{2},s_{0})=&\int_{0}^{s_{0}} \left(
   \frac{s^4}{53760 \pi ^6}-\frac{\dqq m_{s} s^2}{192 \pi ^4}+\frac{13 \dss m_{s} s^2}{1920 \pi ^4}-\frac{\dGG s^2}{2304 \pi ^5}+\frac{\dqq \dss s}{18 \pi ^2}-\frac{\dqq^2 s}{72 \pi ^2}-\frac{\dss^2 s}{72 \pi ^2}-\frac{29 \dqGq m_{s} s}{2304 \pi ^4}\right.\\ &\quad \left.+\frac{29 \dsGs m_{s} s}{2304 \pi ^4}-\frac{5 \dqGq \dqq}{192 \pi ^2}-\frac{5 \dsGs \dss}{192 \pi ^2}+\frac{29 \dqq \dsGs}{576 \pi ^2}+\frac{29 \dqGq \dss}{576 \pi ^2}\right.\\ &\quad \left.+\frac{5 \dGG \dqq m_{s}}{2304 \pi ^3}+\frac{\dGG \dss m_{s}}{2304 \pi ^3}
   \right)e^{-\frac{s}{M_{B}^{2}}} ds 
   -\frac{\dqGq^2}{128 \pi ^2}+\frac{17 \dsGs \dqGq}{576 \pi ^2}+\frac{\dGG m_{s} \dqGq}{256 \pi ^3}\\ &\quad+\frac{2 \dqq \dss^2 m_{s}}{9}+\frac{\dqq^2 \dss m_{s}}{3} +\frac{\dGG \dqq \dss}{108 \pi }-\frac{\dGG \dqq^2}{216 \pi }-\frac{\dGG \dss^2}{216 \pi }-\frac{\dsGs^2}{128 \pi ^2}\\ &\quad-\frac{\dGG \dsGs m_{s}}{6912 \pi ^3}
   \, ,
\end{aligned}
\end{equation}

\begin{equation}
\begin{aligned}
     \Pi_{s}^{2'}(M_{B}^{2},s_{0})=&\int_{0}^{s_{0}} \left(
   \frac{s^4}{26880 \pi ^6}-\frac{\dqq m_{s} s^2}{96 \pi ^4}+\frac{13 \dss m_{s} s^2}{960 \pi ^4}-\frac{\dGG s^2}{1152 \pi ^5}+\frac{\dqq \dss s}{9 \pi ^2}-\frac{\dqq^2 s}{36 \pi ^2}-\frac{\dss^2 s}{36 \pi ^2}-\frac{61 \dqGq m_{s} s}{2304 \pi ^4}\right.\\ &\quad \left.+\frac{55 \dsGs m_{s} s}{2304 \pi ^4}-\frac{3 \dqGq \dqq}{64 \pi ^2}-\frac{3 \dsGs \dss}{64 \pi ^2}+\frac{61 \dqq \dsGs}{576 \pi ^2}+\frac{61 \dqGq \dss}{576 \pi ^2}\right.\\ &\quad \left.-\frac{11 \dGG \dqq m_{s}}{2304 \pi ^3}-\frac{19 \dGG \dss m_{s}}{2304 \pi ^3}
   \right)e^{-\frac{s}{M_{B}^{2}}} ds 
   -\frac{5 \dqGq^2}{384 \pi ^2}+\frac{37 \dsGs \dqGq}{576 \pi ^2}\\ &\quad +\frac{\dGG m_{s} \dqGq}{256 \pi ^3}+\frac{4\dqq \dss^2 m_{s}}{9} +\frac{2\dqq^2 \dss m_{s}}{3} +\frac{5 \dGG \dqq \dss}{108 \pi }+\frac{\dGG \dqq^2}{216 \pi }\\ &\quad +\frac{\dGG \dss^2}{216 \pi }-\frac{5 \dsGs^2}{384 \pi ^2}-\frac{29 \dGG \dsGs m_{s}}{6912 \pi ^3}
   \, ,
\end{aligned}
\end{equation}

For the fully strange $2^{-+}$  $ss\bar{s}\bar{s}$ tetraquark system, the correlation functions after Borel transformation are
\begin{equation}
\begin{aligned}
    \Pi_{ss}^{s1}(M_{B}^{2},s_{0})=&\int_{0}^{s_{0}} \left(
      \frac{s^4}{53760 \pi ^6}+\frac{23 \dss m_{s} s^2}{960 \pi ^4}-\frac{\dGG s^2}{2304 \pi ^5}-\frac{\dss^2 s}{12 \pi ^2}+\frac{29 \dsGs m_{s} s}{576 \pi ^4}-\frac{11 \dsGs \dss}{72 \pi ^2}\right.\\ &\quad \left.-\frac{\dGG \dss m_{s}}{288 \pi ^3}
      \right)e^{-\frac{s}{M_{B}^{2}}} ds +\frac{2 m_{s} \dss^3}{9}-\frac{\dGG \dss^2}{54 \pi }-\frac{13 \dsGs^2}{288 \pi ^2}-\frac{7 \dGG \dsGs m_{s}}{864 \pi ^3}\, .
\end{aligned}
\end{equation}


\begin{thebibliography}{10}

\bibitem{Gell-Mann:1964ewy}
M.~{Gell-Mann}, Phys. Lett. \textbf{8}, 214 (1964)

\bibitem{Zweig:1964jf}
G.~Zweig, D.~B. Lichtenberg and S.~P. Rosen (Editors), {{DEVELOPMENTS IN THE QUARK THEORY OF HADRONS}}. {{VOL}}. 1. 1964 - 1978, 22--101 (1964)

\bibitem{ParticleDataGroup:2022pth}
R.~L. Workman et~al., PTEP \textbf{2022}, 083C01 (2022)

\bibitem{Meyer:2015eta}
C.~A. Meyer and E.~S. Swanson, Prog. Part. Nucl. Phys. \textbf{82}, 21 (2015)

\bibitem{Chen:2016qju}
H.-X. Chen, W.~Chen, X.~Liu, and S.-L. Zhu, Phys. Rep. \textbf{639}, 1 (2016)

\bibitem{Clement:2016vnl}
H.~Clement, Prog. Part. Nucl. Phys. \textbf{93}, 195 (2017)

\bibitem{Guo:2017jvc}
F.-K. Guo, C.~Hanhart, U.-G. Mei{\ss}ner, Q.~Wang, Q.~Zhao, and B.-S. Zou, Rev. Mod. Phys. \textbf{90}, 015004 (2018)

\bibitem{Liu:2019zoy}
Y.-R. Liu, H.-X. Chen, W.~Chen, X.~Liu, and S.-L. Zhu, Prog. Part. Nucl. Phys. \textbf{107}, 237 (2019)

\bibitem{Brambilla:2019esw}
N.~Brambilla et~al., Phys. Rep. \textbf{873}, 1 (2020)

\bibitem{Chen:2022asf}
H.-X. Chen, W.~Chen, X.~Liu, Y.-R. Liu, and S.-L. Zhu, Rept. Prog. Phys. \textbf{86}, 026201 (2023)

\bibitem{Liu:2024uxn}
M.-Z. Liu, Y.-W. Pan, Z.-W. Liu, T.-W. Wu, J.-X. Lu, and L.-S. Geng, Phys. Rept. \textbf{1108}, 2368 (2025)

\bibitem{Meng:2022ozq}
L.~Meng, B.~Wang, G.-J. Wang, and S.-L. Zhu, Phys. Rept. \textbf{1019}, 1 (2023)

\bibitem{Esposito:2016noz}
A.~Esposito, A.~Pilloni, and A.~D. Polosa, Phys. Rept. \textbf{668}, 1 (2017)

\bibitem{Lebed:2016hpi}
R.~F. Lebed, R.~E. Mitchell, and E.~S. Swanson, Prog. Part. Nucl. Phys. \textbf{93}, 143 (2017)

\bibitem{Wang:2025sic}
Z.-G. Wang, arXiv:2502.11351 [hep-ph]  (2025)

\bibitem{BaBar:2006gsq}
B.~Aubert et~al., Phys. Rev. D \textbf{74}, 091103 (2006)

\bibitem{BES:2007sqy}
M.~Ablikim et~al., Phys. Rev. Lett. \textbf{100}, 102003 (2008)

\bibitem{BESIII:2014ybv}
M.~Ablikim et~al., Phys. Rev. D \textbf{91}, 052017 (2015)

\bibitem{BESIII:2017qkh}
M.~Ablikim et~al., Phys. Rev. D \textbf{99}, 012014 (2019)

\bibitem{BaBar:2007ptr}
B.~Aubert et~al., Phys. Rev. D \textbf{76}, 012008 (2007)

\bibitem{BaBar:2007ceh}
B.~Aubert et~al., Phys. Rev. D \textbf{77}, 092002 (2008)

\bibitem{BaBar:2011btv}
J.~P. Lees et~al., Phys. Rev. D \textbf{86}, 012008 (2012)

\bibitem{Belle:2008kuo}
C.~P. Shen et~al., Phys. Rev. D \textbf{80}, 031101 (2009)

\bibitem{BESIII:2018ldc}
M.~Ablikim et~al., Phys. Rev. D \textbf{99}, 032001 (2019)

\bibitem{BESIII:2022iwi}
M.~Ablikim et~al., Phys. Rev. D \textbf{106}, 072012 (2022)

\bibitem{BESIII:2022riz}
M.~Ablikim et~al., Phys. Rev. Lett. \textbf{129}, 192002 (2022)

\bibitem{Deng:2010zzd}
C.~Deng, J.~Ping, F.~Wang, and T.~Goldman, Phys. Rev. D \textbf{82}, 074001 (2010)

\bibitem{Wang:2006ri}
Z.-G. Wang, Nucl. Phys. A \textbf{791}, 106 (2007)

\bibitem{Ke:2018evd}
H.-W. Ke and X.-Q. Li, Phys. Rev. D \textbf{99}, 036014 (2019)

\bibitem{Chen:2018kuu}
H.-X. Chen, C.-P. Shen, and S.-L. Zhu, Phys. Rev. D \textbf{98}, 014011 (2018)

\bibitem{Ho:2019org}
J.~Ho, R.~Berg, T.~G. Steele, W.~Chen, and D.~Harnett, Phys. Rev. D \textbf{100}, 034012 (2019)

\bibitem{Deng:2013aca}
C.~Deng, J.~Ping, Y.~Yang, and F.~Wang, Phys. Rev. D \textbf{88}, 074007 (2013)

\bibitem{Lu:2019ira}
Q.-F. L{\"u}, K.-L. Wang, and Y.-B. Dong, Chin. Phys. C \textbf{44}, 024101 (2020)

\bibitem{Azizi:2019ecm}
K.~Azizi, S.~S. Agaev, and H.~Sundu, Nucl. Phys. B \textbf{948}, 114789 (2019)

\bibitem{Chen:2008ne}
H.-X. Chen, A.~Hosaka, and S.-L. Zhu, Phys. Rev. D \textbf{78}, 117502 (2008)

\bibitem{Chen:2008qw}
H.-X. Chen, A.~Hosaka, and S.-L. Zhu, Phys. Rev. D \textbf{78}, 054017 (2008)

\bibitem{Chen:2022qpd}
H.-X. Chen, N.~Su, and S.-L. Zhu, Chin. Phys. Lett. \textbf{39}, 051201 (2022)

\bibitem{Qiu:2022ktc}
L.~Qiu and Q.~Zhao, Chin. Phys. C \textbf{46}, 051001 (2022)

\bibitem{Shastry:2022upd}
V.~Shastry, PoS \textbf{ICHEP2022}, 779 (2022)

\bibitem{Chen:2023ukh}
B.~Chen, S.-Q. Luo, and X.~Liu, arXiv:2302.06785  (2023)

\bibitem{Chen:2022isv}
F.~Chen et~al., Phys. Rev. D \textbf{107}, 054511 (2023)

\bibitem{Shastry:2023ths}
V.~Shastry and F.~Giacosa, Nucl. Phys. A \textbf{1037}, 122683 (2023)

\bibitem{Wan:2022xkx}
B.-D. Wan, S.-Q. Zhang, and C.-F. Qiao, Phys. Rev. D \textbf{106}, 074003 (2022)

\bibitem{Yang:2022rck}
F.~Yang, H.~Q. Zhu, and Y.~Huang, Nucl. Phys. A \textbf{1030}, 122571 (2023)

\bibitem{Jiao:2009ra}
C.-K. Jiao, W.~Chen, H.-X. Chen, and S.-L. Zhu, Phys. Rev. D \textbf{79}, 114034 (2009)

\bibitem{Huang:2016rro}
Z.-R. Huang, W.~Chen, T.~G. Steele, Z.-F. Zhang, and H.-Y. Jin, Phys. Rev. D \textbf{95}, 076017 (2017)

\bibitem{Du:2012pn}
M.-L. Du, W.~Chen, X.-L. Chen, and S.-L. Zhu, Chin. Phys. C \textbf{37}, 033104 (2013)

\bibitem{Fu:2018ngx}
Y.-C. Fu, Z.-R. Huang, Z.-F. Zhang, and W.~Chen, Phys. Rev. D \textbf{99}, 014025 (2019)

\bibitem{Ray:2022fcl}
K.~Ray, D.~Harnett, and T.~G. Steele, Phys. Rev. D \textbf{108}, 034001 (2023)

\bibitem{Xi:2023byo}
H.-Z. Xi, Y.-W. Jiang, H.-X. Chen, A.~Hosaka, and N.~Su, Phys. Rev. D \textbf{108}, 094019 (2023)

\bibitem{Lodha:2024bwn}
C.~Lodha and A.~K. Rai, Few Body Syst. \textbf{65}, 99 (2024)

\bibitem{Su:2022eun}
N.~Su and H.-X. Chen, Phys. Rev. D \textbf{106}, 014023 (2022)

\bibitem{Wang:2019nln}
Z.-G. Wang, Adv. High Energy Phys. \textbf{2020}, 6438730 (2020)

\bibitem{Cui:2019roq}
E.-L. Cui, H.-M. Yang, H.-X. Chen, W.~Chen, and C.-P. Shen, Eur. Phys. J. C \textbf{79}, 232 (2019)

\bibitem{BESIII:2010gmv}
M.~Ablikim et~al., Phys. Rev. Lett. \textbf{106}, 072002 (2011)

\bibitem{BESIII:2019wkp}
M.~Ablikim et~al., Eur. Phys. J. C \textbf{80}, 746 (2020)

\bibitem{BESIII:2023wfi}
M.~Ablikim et~al., Phys. Rev. Lett. \textbf{132}, 181901 (2024)

\bibitem{Li:2025bko}
X.-T. Li, G.-Y. Wang, and Q.-A. Zhang, arXiv:2501.15224 [hep-ph]  (2025)

\bibitem{Li:2024fko}
H.-n. Li, Chin. Phys. Lett. \textbf{41}, 101101 (2024)

\bibitem{Cao:2024mfn}
J.~Cao et~al., Phys. Rev. D \textbf{110}, 054046 (2024)

\bibitem{Yu:2011ta}
J.-S. Yu, Z.-F. Sun, X.~Liu, and Q.~Zhao, Phys. Rev. D \textbf{83}, 114007 (2011)

\bibitem{BESIIICollaboration:2022kwh}
M.~Ablikim et~al., Phys. Rev. Lett. \textbf{129}, 042001 (2022)

\bibitem{Zhang:2022obn}
S.-Q. Zhang, B.-D. Wan, L.~Tang, and C.-F. Qiao, Phys. Rev. D \textbf{106}, 074010 (2022)

\bibitem{Wang:2024yvo}
L.-M. Wang, W.-X. Tian, T.-Y. Li, C.-X. Liu, and X.~Liu, arXiv:2408.05908 [hep-ph]  (2024)

\bibitem{Lodha:2024yfn}
C.~Lodha and A.~K. Rai, arXiv:2412.05874 [hep-ph]  (2024)

\bibitem{Chen:2010ze}
W.~Chen and S.-L. Zhu, Phys. Rev. D \textbf{83}, 034010 (2011)

\bibitem{Wang:2020cme}
Z.-G. Wang, Phys. Rev. D \textbf{101}, 074011 (2020)

\bibitem{Jamin:2002ev}
M.~Jamin, Phys. Lett. B \textbf{538}, 71 (2002)

\bibitem{Narison:2011xe}
S.~Narison, Phys. Lett. B \textbf{706}, 412 (2012)

\bibitem{Narison:2018dcr}
S.~Narison, Int. J. Mod. Phys. A \textbf{33}, 1850045 (2018)

\end{thebibliography}

\end{document}